# A matheuristic algorithm for the single-source capacitated facility location problem and its variants

Yunfeng Kong

Key Laboratory of Geospatial Technology for the Middle and Lower Yellow River Regions, Ministry of Education, Henan University, Kaifang, 475000 China

**Abstract:** This article presents a matheuristic algorithm for the single-source capacitated facility location problem (SSCFLP) and its variants: SSCFLP with $K$ facilities (SSCKFLP), SSCFLP with contiguous service areas (CFLSAP), and SSCFLP with $K$ facilities and contiguous service areas (CKFLSAP). The algorithm starts from an initial solution, and iteratively improves the solution by exactly solving large neighborhood-based sub-problems. The performance of the algorithm is tested on 5 sets of SSCFLP benchmark instances. Among the 272 instances, 191 optimal solutions are found, and 35 best-known solutions are updated. For the largest set of instances with 300-1000 facilities and 300-1500 customers (Avella and Boccia 2009), the proposed algorithm outperforms existing methods in terms of the solution quality and the computational time. Furthermore, based on two geographic areas, two sets of instances are generated to test the algorithm for solving SSCFLP and its variants. The solutions found by the proposed algorithm approximate optimal solutions or the lower bounds with average gaps of 0.07% for SSCFLP, 0.22% for CFLSAP, 0.04% for SSCKFLP, and 0.13% for CKFLSAP.

## 1 Introduction

Facility location problems aim to investigate where to optimally locate a set of facilities. They have been widely used in both public and private facility planning, such as schools, healthcare centers, disaster shelters, warehouses, and logistic centers. The problems can be classified according to application specifications such as continuous or discrete locations for setting facilities, capacitated or uncapacitated facilities, assigning each customer to single or multiple facilities, and how to define the decision objective. For problem definition, mathematical formulation, algorithm design and real world applications of various location problems, read the book Location Science edited by Laporte et al. (2015).

The single-source capacitated facility location problem (SSCFLP), one of the most difficult location problems, has been extensively discussed since 1980s. Let $I$ be a set of candidate locations for opening facilities, and $J$ be a set of customers. Each facility at location $i$ has a fixed opening cost $f_i$ and a service capacity $s_i$. Each client $j$ has a demand $d_j$ that must be served by a single facility. The cost for satisfying the demand of customer $j$ from a facility located at $i$ is $c_{ij}$. The SSCFLP can be formulated as follows.

$$\text{Minimize} \sum_{i\in I} f_i + \sum_{i\in I}\sum_{j\in J} c_{ij}x_{ij} \quad (1)$$
$$\text{Subject to} \sum_{i\in I} x_{ij} = 1, \forall j \in J \quad (2)$$
$$\sum_{j\in J} d_j x_{ij} \le s_i y_i, \forall i \in I \quad (3)$$
$$y_i = \{0,1\}, \forall i \in I \quad (4)$$
$$x_{ij} = \{0,1\}, \forall i \in I, j \in J \quad (5)$$

The binary variables $y_i$ in constraints (4) indicate whether the candidate facility is opened at location $i$, and the binary variables $x_{ij}$ in constraints (5) indicate whether the demand of customer $j$ is served by the facility at location $i$. The objective function (1) minimizes the total cost of opening facilities and the total cost of assigning customers to open facilities. The constraints (2) ensure each customer is served by a single facility. The constraints (3) confirm that the customers must be assigned to open facilities and that the total demand assigned to a facility cannot exceed its maximum capacity.

There are two general approaches to solve SSCFLP since 1980s: exact and heuristics (Basu et al. 2015; Ulukan & Demircioğlu 2015). Exact methods include branch and bound (Neebe & Rao 1983; Holmberg et al. 1999), branch-and-price (Díaz & Fernández 2002), and cutting plane (Avella & Boccia 2009; Yang et al. 2012; Gadegaard et al. 2018), or CPLEX branch-and-cut (Yang et al. 2012; Guastaroba & Speranza 2014; Caserta & Voß 2020). Some sets of well-known SSCFLP benchmark instances can be solved optimally and efficiently by these exact methods (Holmberg et al. 1999; Díaz & Fernández 2002). The benchmark instances with 30-80 facilities and 200-400 customers were also successfully solved by cut-and-solve method (Yang et al. 2012) and improved cut-and-solve method (Gadegaard et al. 2018). For the 100 large instances with 300-1000 facilities and 300-1500 customers (Avella and Boccia 2009), 45 instances were optimally solved by CPLEX branch-and-cut algorithm (Guastaroba & Speranza 2014; Caserta & Voß 2020). However, since SSCFLP is nondeterministic polynomial time hard (NP-hard) in strong sense, it is challenging to efficiently solve large SSCFLP instances by exact methods.

There are various heuristic algorithms for the SSCFLP. Lagrangian relaxation-based heuristic (LH) has been extensively investigated since 1980s (Barcelo & Casanova 1984; Klincewicz & Luss 1986; Beasley 1993; Sridharan 1993; Agar & Salhi 1998; Hindi & Pienkosz 1999; Rönnqvist et al. 1999; Cortinhal & Captivo 2003; Oliveira et al. 2020). Based on the dual models that relaxes the capacity constraints, and/or the assignment constraints, there methods repeatedly perform the following procedures: (1) solve the dual model, and update the lower bound; (2) find a feasible solution using the dual model solution, and update the upper bound; and (3) update the Lagrangian multipliers using the gradient descent method. Various LH techniques for solving facility location problems were surveyed in Galvão & Marianov (2011). Since LH is simple and fast, it is usually used to generate initial solutions for many metaheuristics. It can also find a tight lower bound on SSCFLP, and thus is useful for evaluating the solution quality. Other heuristic methods for SSCFLP include tabu search (Filho & Galvao1998; Delmaire et al. 1999; Cortinhal & Captivo 2003), very large neighborhood search (Ahuja et al. 2004; Tran et al. 2017), scatter search(Contreras & Diaz, 2008), ant colony system (Chen & Ting 2008), kernel search (Guastaroba & Speranza 2014), and corridor method (Caserta & Voß, 2020).

In the last 10 years, the performance of solving large SSCFLP, in terms of solution quality and computational time, has been progressively increased by algorithms such as kernel search (Guastaroba & Speranza 2014), multi-exchange heuristic (Tran et al. 2017), and corridor method (Caserta & Voß 2020). The largest set of instances with 300-1000 facilities and 300-1500 customers (Avella & Boccia 2009) were solved by the three algorithms with average solution gaps 0.64% (Guastaroba & Speranza 2014), 0.60% (Tran et al. 2017) and 0.50% (Caserta & Voß 2020). The kernel search method is designed to exactly solve a sequence of subproblems, each of which is restricted to a subset of the decision variables. The subsets of decision variables are constructed using the optimal values of the linear relaxation. The multi-exchange heuristic explores very large neighborhoods based on dynamically-built improvement hypergraphs. The corridor method exploits Lagrangean relaxation solutions and builds corridors by introducing constraints around the incumbent solution, which limits the size of the solution space explored at each iteration. However, since SSCFLP is NP-hard in strong sense, it is difficult to solve large instances in a reasonable computation time. For example, the instances with 700 facilities and 700 customers in Avella & Boccia (2009), were solved in 7747, 5244, 4992 and 912 seconds by CPLEX, kernel search, multi-exchange heuristic and corridor method, respectively.

In real-world service location planning, more criteria are required. The first criterion is the quantity constraints on the facilities (Aardal et al. 2015; Wang 2017). Second, the contiguity of facility service areas are frequently required in some facility site selection applications. Service districting is one of the most important issues associated with the provision of some public services such as homecare and compulsory education (Benzarti et al. 2013; Kalcsics 2015; Kong et al. 2017; Wang & Kong 2021). For example, the service areas for compulsory schools in urban China are usually continuous, so as to avoid some enrollment controversies. The healthcare centers in China are also requested to serve the residents living in a predefined area with explicit boundary. Thereafter, it is necessary to investigate the SSCFLP with additional criteria, especially with contiguous facility service areas.

In this article, a matheuristic algorithm is proposed for SSCFLP and its variants. The algorithm starts from an initial solution, and then iteratively improves the solution by searching large neighborhood of current solution. The performance of the algorithm was tested on five sets of SSCFLP benchmark instances. Experimentation shows that the matheuristic algorithm outperforms the existing methods. Among the 272 SSCFLP instances, 191 optimal solutions are found, and 35 best-known solution are updated. The algorithm was also used to solve three variants of SSCFLP: SSCFLP with $K$ facilities and/or with connective service areas. The solutions of variant problems approximate optimal solutions or the lower bounds with average gaps less than 0.22%.

There are three contributions in this article. First, the model of SSCFLP with contiguous facility service areas is mathematically formulated, and is verified by solving two sets of instances. Second, a simple but effective matheuristic algorithm is proposed for SSCFLP and its variants. Third, for the largest set of SSCFLP instances with 300-1000 facilities and 300-1500 customers (Avella & Boccia 2009), 31 best-known solutions are updated.

The article is organized as follows. Section 2 defines three variants of SSCFLP. Section 3 describes the matheuristic algorithm. Section 4 reports the solution results from well-known benchmark instances and newly-generated instances. Section 5 gives conclusion remarks.

# 2 SSCFLP variants

Let $J$ be a set of spatial units in a geographical area, and each unit $j$ has service demand $d_j$. Let set $I$, a subset of $J$ ($I \subseteq J$), be candidate locations for setting facilities, and unit $i$ has maximum capacity $s_i$. Let $c_{ij}$ be the cost of satisfying the demand of customer $j$ from a facility located in unit $i$. The model formulations (1)-(5) can be used to solve the SSCFLP instances associated with geographical areas.

It is possible to extend the SSCFLP model by adding contiguity constraints on facility service areas. Three types of constraints on contiguity criterion, tree-based, order-based and flow-based, were proposed for the p-Regions problem in Duque et al. (2011). In the flow model, the service area contiguity is ensured by establishing a flow route from each spatial unit to its facility unit within the facility service area. The flow model has been adaptively formulated for service area problem (Wang & Kong 2021; Kong 2021) and districting problem (Kong et al. 2019; Kong 2021). It is also feasible to be embedded in SSCFLP. Let $a_{jk}$ indicate whether unit $j$ and $k$ share a border, and $N_j$ be a set of units that are adjacent to unit $j$ ($N_j = \{k|a_{jk} = 1\}$). Let $f_{ijk}$ be decision variables that indicate the flow volume from unit $j$ to unit $k$ in service area $i$, the flow model for SSCFLP can be formulated as follows:

$$f_{ijk} \le n * x_{ij}, \forall i \in I, j \in J, k \in N_j \quad (6)$$
$$f_{ijk} \le n * x_{ik}, \forall i \in I, j \in J, k \in N_j \quad (7)$$
$$\sum_{k \in N_j} f_{ijk} - \sum_{k \in N_j} f_{ikj} \ge x_{ij}, \forall i \in I, j \in J \backslash i \quad (8)$$
$$f_{ijk} \ge 0, \forall i \in I, j \in J, k \in N_j \quad (9)$$

Since the optimal number of facilities is unknown in prior, the maximum flow volume does not exceed $|J| - 1$. Constraints (6) and (7) ensure that flows can only be created within a service area. Constraints (6) state that if unit $j$ is not serviced by facility $i$ ($x_{ij} = 0$), these is no any outflow from unit $j$ to its neighbors; otherwise, these is an outflow with maximum volume $n$ from unit $j$ to its neighbors ($n$=$|J| - 1$). Constraints (7) state that if unit $k$ is not serviced by facility $i$ ($x_{ik} = 0$), these is no any inflow from its neighbors to unit $k$; otherwise, these is an inflow with maximum volume $n$ from its neighbors to unit $k$. Constraints (8) guarantee that if customer unit $j$ is served by facility $i$ ($x_{ij} = 1$), one-unit flow will be created in unit $j$, combined with the inflows, and runs off from unit $j$. Since the facility unit serves as the sink unit of its service area and there is no any outflow from it, the candidate locations is excluded from its service area in constraints (8).

It is easy to extend SSCFLP model by adding the quantity constraints on the facilities, such as:

$$\sum_{i \in I} y_i = K \quad (10)$$
$$K_{min} \le \sum_{i \in I} y_i \le K_{max} \quad (11)$$

Based on the formulations above, three variant problems can be defined:

(1) SSCFLP with $K$ facilities (SSCKFLP): objective function (1) subject to (2)-(5) and (10);

(2) SSCFLP with contiguous service areas (CFLSAP): objective function (1) subject to (2)-(9);

(3) SSCFLP with $K$ facilities and contiguous service areas (CKFLSAP): objective function (1) subject to (2)-(10).

# 3 Matheuristic algorithm

A matheuristic algorithm is designed for solving SSCFLP and its variants. The availability of the state-of-the-art MIP solvers, such as IBM CPLEX Oprimizor and Gurobi Optimizer, has created new opportunities in the design of matheuristics that combine heuristic schemes with mixed integer linear programming strategies (Archetti et al. 2104; Kergosien et al. 2021). Matheuristics have been shown to be quite effective in solving complex MIP problems (Maniezzo et al. 2010). A matheuristic was also used to efficiently solve large-size p-median problem instances (Gnägi & Baumann, 2021).

The idea of the matheuristic algorithm for facility location problem is simple. It starts from an initial solution, and improves the current solution progressively by exactly solving large neighborhood-based sub-problems. The algorithm is outlined as follows.

Algorithm: Matheuristic for SSCFLP and its variants

Parameter: number of consecutive loops that the best solution is not updated (*mloops*).

1. $s$=GenerateInitialSolution();
2. *notImpr*=0;
3. WHILE *notImpr* < *mloops*:
4. $I^*, J^*$=SelectNeighborhood($s$);
5. $s^*$=SolveSubProblem($I^*, J^*$);
6. $s'$=CreateNewSolution($s$, $s^*$);
7. IF problem is CFLSAP or CKFLSAP:
8. $s'$=RepairAndSearch($s'$);
9. IF $f(s')<f(s)$: $s = s'$, *notImpr*=0;
10. ELSE: *notImpr*+=1;
11. Output($s$).

In the algorithm, an initial solution is generated in step (1). There are multiple methods to create an initial solution, such as Lagrangian relaxation-based heuristic (Holmberg et al. 1999), linear relaxation-based heuristic, and simple construction method. For some instances, it might be difficult to generate a feasible solution. One possible way is to use soft constraints on facility capacities. Let decision variable $H_i$ ($H_i \geq 0, \forall i \in I$) be the service overload of facility $i$, the capacity constraints (3) and objective function (1) can be replaced by formulations (12) and (13), respectively. Constraints (12) are the soft constraints on maximum service capacities. The service overloads $H_i$ are penalized by multiplying a large enough coefficient α in objective function (13). In case of the penalty cost is reduced to zero, the constraints (3) are satisfied and the solution is feasible to the original problem.

$$\sum_{j \in J} d_i x_{ij} \leq s_i y_i + H_i, \forall i \in I \quad (12)$$

$$Min. \sum_{i \in I} f_i + \sum_{i \in I} \sum_{j \in J} c_{ij} x_{ij} + \alpha \sum_{i \in I} H_i \quad (13)$$

Stating from an initial solution, the algorithm will iteratively improve the current solution by the following procedures: select a large neighborhood randomly from current solution in step (4); solve the neighborhood-based sub-problem exactly in step (5); create a new solution by combining current solution and the sub-problem solution in step (6); repair and search current solution for CFLSAP or CKFLSAP instance in step (8); and update current solution in step (9). The iterations will be terminated in case of the best solution is not updated in *mloops* consecutive loops.

In step (4), the neighborhood is defined by choosing a subset of facility locations and a subset of customers that are spatially clustered, denoted as $I^*(I^* \subset I)$ and $J^*(J^* \subset J)$, respectively. The subsets are prepared in three steps. First, select a customer randomly from all customers, and then choose $Q$ nearest open facilities to the customer, denoted as set $I'$. Second, choose all the customers that assigned to the $Q$ facilities, denoted as $J^*$. Third, choose the nearest candidate location to each customer in $J^*$, denoted as set $I''$, and let facility location set $I^* = I' \cup I''$. In case of the size of set $I^*$ is too large, part locations need to be deleted randomly from set $I''$ for ensuring that $|I^*| * |J^*| < U_{max}$, and thus the related sub-problem can be solved efficiently by a MIP solver. In this article, the parameter $Q$ is selected uniformly at random in the range [$Q_{min}$, $Q_{max}$]. Let $L$ be the number of open facilities in current solution, $Q_{min}$=Min($L$/2, 7) and $Q_{max}$=Min($L$, 10). Consequently, Q=$L$/2~$L$ ($L$≤10), $Q$= $L$/2~10 (11≤$L$≤13), or $Q$=7~10 ($L$≥14). The second parameter $U_{max}$ is set to Min(3000, $|I| * |J|/10$).

In step (5), a SSCFLP/SSCKFLP model is built by using facility set $I^*$ and customer set $J^*$, and then solved by a MIP solver. Since the neighborhood size is limited by the parameters $Q$ and $U_{max}$, the model can be efficiently solved. In step (6), a new solution $s'$ is created by deleting facilities $I'$ and customers $J^*$ from current solution $s$, inserting the open facilities in sub-solution solution $s^*$ into solution $s$, and then assigning the customers $J^*$ to facilities according to the sub-solution solution $s^*$. Note that SSCFLP model is used for SSCFLP or CFLSAP instance, and SSCKFLP model is used for SSCKFLP or CKFLSAP instance.

For CFLSAP or SSCKFLP instance, it is necessary to repair the solution such that the facility service areas are contiguous. The solution is repaired as follows: find the fragmented units in current solution; delete these units from current solution; and insert each deleted unit to one of its neighboring service area in a greedy manner. The repaired solution will become worse and even with service overload. A local search procedure can usually improve the solution. In step (8), two local search operators are used to improve the solutions: one-unit shift and two-unit shift (Butsch et al. 2014; Kong et al. 2017; Kong 2021). The local search operators attempted to move one or two units located on the boundary to their neighboring service areas. Note that only the feasible moves are allowed, because when moving a boundary unit from its original area to a destination area, the original area may be non-contiguous.

If the new solution is better that the current solution, it will be used to replace the current solution, shown in step (10). Function $f(s)$ is the cost objective of solution $s$.

The proposed algorithm was implemented by using the Python programming language. In Python script, the PuLP, a linear programming toolkit (https://github.com/coin-or/pulp), is used to generate sub-problem models, and solve the models by calling IBM ILOG CPLEX Optimizer 12.6 (https://www.ibm.com/products/ilog-cplex-optimization-studio). The algorithm code can be downloaded from webpage https://github.com/yfkong/Unified.

# 4 Experiment

### 4.1 Benchmark instances of SSCFLP

The algorithm's performance is tested using five SSCFLP benchmark datasets with 10-1000 facilities and 50-1500 customers. The dataset name, source, number of instances and instance sizes for each dataset are shown in Table 1. Datasets OR-Lib (Ahuja et al. 2004), Holmberg (Holmberg et al. 1999), Yang (Yang et al. 2012) and Tebd1 (Avella & Boccia 2009) can be downloaded from webpage https://or-brescia.unibs.it/instances/instances_sscflp. Dataset TB4 (Gadegaard et al. 2018) can be downloaded from webpage https://github.com/SuneGadegaard/SSCFLPsolver. In addition, each dataset are classified into several groups according to the instance size.

**Table 1 SSCFLP benchmark instances**

| Dataset | Instance group | Num. of instances | $\|I\|$ | $\|J\|$ |
|---|---|---|---|---|
| OR-Lib (Ahuja et al. 2004) | OR1 (cap61-cap74) | 8 | 16 | 50 |
| | OR2 (cap91-cap104) | 8 | 25 | 50 |
| | OR3 (cap121-cap134) | 8 | 50 | 50 |
| | OR4 (capax, capbx, capcx) | 12 | 100 | 1000 |
| Holmberg (Holmberg et al. 1999) | H1 (p1-p12) | 12 | 10 | 50 |
| | H2 (p13-p24) | 12 | 20 | 50 |
| | H3 (p25-p40) | 16 | 30 | 150 |
| | H4 (p41-p55) | 15 | 10-30 | 70-100 |
| | H5 (p56-p71) | 16 | 30 | 200 |
| Yang (Yang et al. 2012) | Y1 (30_200_x) | 5 | 30 | 200 |
| | Y2 (60_200_x) | 5 | 60 | 200 |
| | Y3 (60_300_x) | 5 | 60 | 300 |
| | Y4 (80_400_x) | 5 | 80 | 400 |
| TB4 (Gadegaard et al. 2018) | G1 (50_100_x_x) | 15 | 50 | 100 |
| | G2 (50_200_x_x) | 15 | 50 | 200 |
| | G3 (60_300_x_x) | 15 | 60 | 300 |
| Tebd1 (Avella and Boccia 2009) | T1 (i300_x) | 20 | 300 | 300 |
| | T2 (i3001500_x) | 20 | 300 | 1500 |
| | T3 (i500_x) | 20 | 500 | 500 |
| | T4 (i700_x) | 20 | 700 | 700 |
| | T5 (i1000_x) | 20 | 1000 | 1000 |

Each instance was repeatedly solved for five times. The algorithm parameter *mloops* was set as 10 for OR1, OR2, OR3, H1 and H2 instances, 20 for H3, H4 and H5 instances, 50 for OR4 and T2 instances, and 100 for others. Since the initial solution is generated randomly, different solutions will be obtained by repeatedly executing the algorithm. The detailed solutions for all instances are shown in the appendix file of this article.

All the computational results in this article were obtained from a desktop computer with Intel Core I7-6700 CPU 3.40 GHz, 8 GB RAM and the Windows 10 operating system. The Python script runs in PyPy 6.0, a fast and compliant implementation of the Python language (see http://pypy.org), in order to speed up the algorithm.

To verify the optimality of the solutions, the lower bound and upper bound of the objective for each instance were collected from existing literatures, or found by CPLEX optimizer. Among the 272 instances, 220 solutions are optimal.

Solution results obtained by exact methods, the proposed algorithm and several state-of-the-art heuristics are summarized in Table 2. For exact methods, the number of optimal solutions shown in column #opt was found by CPLEX branch-and-cut method or the improved cut-and-solve method (Gadegaard et al. 2018). The OR1, OR2 and OR3 instances were solved by the author using CPLEX 12.6; the OR4 instances were solved by CPLEX 12.2 (Guastaroba et al. 2014); the Holmberg, Yang and TB4 instances were solved by the improved cut-and-solve method (Gadegaard et al. 2018); and 55 Tbed1 instances were solved optimally by CPLEX 12.6, 45 of them from Caserta & Voß (2020) and 10 from the author. It is found that exact methods can be used to solve small-size instances in datasets OR-Lib and Holmberg. However, it is hard to solve most medium-size and large-size instances in datasets Yang, TB4 and Tbed1. For example, 14 of 20 Yang instances cannot be exactly solved by CPLEX branch-and-cut method within 50000 seconds (Yang et al. 2012). The improved cut-and-solved method is much more efficient than cut-and-solved and CPLEX branch-and-cut, but it is still time-consuming for solving instances in dataset TB4. Consequently, it is necessary to design heuristic algorithm to solve SSCFLP.

In table 2, the average solution gap and computation time for each heuristic method are shown in column Gap and Time, respectively. MH stands for the proposed matheuristic algorithm in this article; KS, HMEH and CM denote the kernel search (Guastaroba & Speranza 2014), hypergraph based multi-exchange heuristic (Tran et al. 2017), and corridor method (Caserta & Voß 2020), respectively. The solution gap is calculated by the formula (14) for the instance that its optimal solution is known, or by formula (15) for the instance that its optimal solution is not found.

$$\text{Gap} = \frac{\text{objective – optimal objective}}{\text{optimal objective}} * 100\% \qquad (14)$$

$$\text{Gap} = \frac{\text{objective – lower bound}}{\text{lower bound}} * 100\% \qquad (15)$$

**Table 2 Summary of SSCFLP solution results**

| Dataset | Group | Exact | | MH | | KS | | HMEH | | CM | |
|---|---|---|---|---|---|---|---|---|---|---|---|
| | | #opt | Time/s | Gap/% | Time/s | Gap/% | Time/s | Gap/% | Time/s | Gap/% | Time /s |
| OR-Lib | OR1 | 8/8 | 0.05 | 0.00 | 1.22 | 0.00 | 0.29 | - | - | - | - |
| | OR2 | 8/8 | 0.06 | 0.00 | 2.23 | 0.00 | 0.39 | - | - | - | - |
| | OR3 | 8/8 | 0.08 | 0.00 | 1.56 | 0.00 | 0.62 | - | - | - | - |
| | OR4 | 12/12 | 112.44 | 0.01 | 149.06 | 0.00 | 34.67 | 0.04 | 42.67 | 0.00 | 43 |
| Holm. | H1 | 12/12 | 0.20 | 0.00 | 1.13 | 0.00 | 0.32 | 0.00 | 0.42 | - | - |
| | H2 | 12/12 | 0.34 | 0.02 | 2.13 | 0.00 | 0.38 | | | - | - |
| | H3 | 16/16 | 2.61 | 0.00 | 8.64 | 0.00 | 2.43 | 0.00 | 4.08 | - | - |
| | H4 | 15/15 | 0.67 | 0.00 | 3.88 | 0.00 | 0.54 | 0.00 | 1.08 | - | - |
| | H5 | 16/16 | 5.29 | 0.00 | 15.59 | 0.00 | 2.32 | 0.00 | 15.53 | - | - |
| Yang | Y1 | 5/5 | 51.00 | 0.00 | 76.31 | 0.00 | 411.28 | - | - | - | - |
| | Y2 | 5/5 | 1261.82 | 0.01 | 27.78 | 0.00 | 1640.42 | - | - | - | - |
| | Y3 | 5/5 | 65.63 | 0.04 | 120.58 | 0.00 | 597.06 | - | - | - | - |
| | Y4 | 5/5 | 228.01 | 0.09 | 232.79 | 0.00 | 1409.11 | - | - | - | - |
| TB4 | G1 | 15/15 | 676 | 0.03 | 86.55 | - | - | - | - | - | - |
| | G2 | 13/15 | 4036 | 0.07 | 47.83 | - | - | - | - | - | - |
| | G3 | 10/15 | 14617 | 0.07 | 61.59 | - | - | - | - | - | - |
| Tebd1 | T1 | 15/20 | 3722 | 0.15 | 60.88 | 0.56 | 2206.96 | 0.54 | 428.03 | 0.23 | 807 |
| | T2 | 20/20 | 47 | 0.00 | 87.12 | 0.00 | 334.71 | 0.01 | 1159.33 | 0.00 | 29 |
| | T3 | 14/20 | 2017 | 0.27 | 158.69 | 0.66 | 4190.28 | 0.52 | 2982.72 | 0.36 | 1024 |
| | T4 | 6/20 | 7744 | 0.47 | 341.18 | 0.90 | 5244.69 | 0.82 | 4992.14 | 0.78 | 912 |
| | T5 | 0/20 | 8275 | 0.58 | 345.58 | 1.07 | 6533.15 | 1.10 | 8582.74 | 1.11 | 932 |

Table 2 shows that different methods for SSCFLP perform very differently on different benchmark datasets. For small-size instances in OR-Lib and Holmberg, exact method is better than heuristics. For medium-size instances in Yang and TB4, the matheuristic is better than kernel search, solve-and-cut, and improved solve-and-cut in terms of solution quality and computation time. For the largest Tbed1 instances with 300-1000 facilities and 300-1500 customers, the matheuristic not only improves the solution quality with lowest gaps, but also reduces the computation time significantly. Note that the times cited in Table 2 cannot be directly compared, since different computers were used in different experiments.

Table 3 shows the number of optimal solutions found by deferent exact and heuristic methods. The optimal solutions were collected in multiple sources. Some incorrect solutions were deleted from this table. The columns MH, CS, CS2, KS, HMEH, CM denote matheuristic, cut-and-solve (Yang et al. 2012), improved cut-and-solve (Gadegaard et al. 2018), kernel search(Guastaroba &

Speranza 2014), hypergraph based multi-exchange (Tran et al. 2017 ), and corridor method (Caserta & Voß 2020), respectively. Since the optimal objectives for instances 50-200-2-4, 50-200-2-5 and 60-200-2-5 are incorrect in Gadegaard et al. (2018), the total number of optimal solutions for dataset TB4 is different than that in Gadegaard et al. (2018).

**Table 3 Optimal SSCFLP solutions obtained by different solution methods**

| Dataset | Group | #ins. | #opt | CPLEX | MH | CS | CS2 | KS | HMEH | CM |
|---|---|---|---|---|---|---|---|---|---|---|
| OR-Lib | OR4 | 12 | 12 | 12 | 11 | - | - | 12 | - | 12 |
| Yang | Y1 | 5 | 5 | 3 | 5 | 5 | 5 | 5 | - | - |
| | Y2 | 5 | 5 | 1 | 4 | 5 | 5 | 4 | - | - |
| | Y3 | 5 | 5 | 1 | 5 | 5 | 5 | 4 | - | - |
| | Y4 | 5 | 5 | 1 | 4 | 5 | 5 | 5 | - | - |
| TB4 | G1 | 15 | 15 | - | 12 | - | 15 | - | - | - |
| | G2 | 15 | 13 | - | 7 | - | 13 | - | - | - |
| | G3 | 15 | 10 | - | 7 | - | 10 | - | - | - |
| Tebd1 | T1 | 20 | 15 | 15 | 12 | - | - | 12 | 10 | 12 |
| | T2 | 20 | 20 | 20 | 17 | - | - | 20 | 14 | 20 |
| | T3 | 20 | 14 | 14 | 7 | - | - | 6 | 4 | 8 |
| | T4 | 20 | 6 | 6 | 5 | - | - | 2 | 0 | 4 |
| | T5 | 20 | 0 | 0 | 0 | - | - | 0 | 0 | 0 |

Among 53 instances in dataset TB4 and Tbed1 that were not solved optimally, 35 best known solutions were updated by the matheuristic algorithm. The best known solutions for Tbed1 instances have been updated progressively by kernel search (Guastaroba & Speranza 2014), multi-exchange heuristic (Tran et al. 2017), corridor method (Caserta & Voß 2020) and the proposed algorithm in this article. The new best objective values are listed in the appendix file, and the detailed solutions can be downloaded from webpage https://github.com/yfkong/Unified. For some instances, such as 60-300-2-2, 60-300-2-4, i700_1, i700_2, i700_3, i700_4, i1000_1, i1000_2, i1000_3, i1000_4, i1000_5, i1000_9, their best objective values were significantly reduced with a mean decrease of 1.03%, ranging between 0.50% and 3.43%.

### 4.2 New instances of SSCFLP and its variants

In order to test facility location problems with contiguous facility service areas, new instances with geographical information are necessary. In this article, two typical geographical regions, ZY and GY, were used to generate instances for SSCFLP and its variants. The urban region ZY has an area of 13.4 square kilometers, consists of 324 spatial units. There are 15 primary schools and 3783 school students in region ZY. The rural region GY has an area of about 1000 square kilometers, consists of 1276 spatial units. There are 18 townships and 819812 residents in region GY.

The number of students (or the number of residents) in each spatial unit is assumed to be the quantity of demand. The schools (or township centers) are supposed to be the candidate facility locations, and the number of students in each school (or the number of residents in each township) is assumed to be its service capacity. In addition, more units are manually selected as candidate facility locations; their capacities are randomly set to be a number between the minimum and maximum capacities of schools (or township centers). Consequently, 36 and 33 units in the two regions were selected to be the candidate units with total supplies of 9195 and 1324763, respectively. The spatial distributions of service demand and supply are illustrated in Figure 1. The grey circles represent the demand quantities and the star symbols represent the candidate facilities.

Two sets of geographical instances were prepared for SSCFLP and its variants. Let the center point of each spatial unit be the customer location and the facility location for service-supply unit. The cost $c_{ij}$ is defined by the Euclidean distance ($d_{ij}$) between facility location $i$ and customer location $j$: $c_{ij} = 1.0 * d_{ij} * d_j$. The fixed cost for each facility is approximately proportional to its maximum capacity: $f_i = (\mu + \varepsilon_i) * s_i$. For region ZY, let $\mu$=0.8 and $\varepsilon_i$= [-0.1, 0.1]; for region GY, let $\mu$=1.8 and $\varepsilon_i$= [-0.2, 0.2]. At this time, two basic instances can be generated.

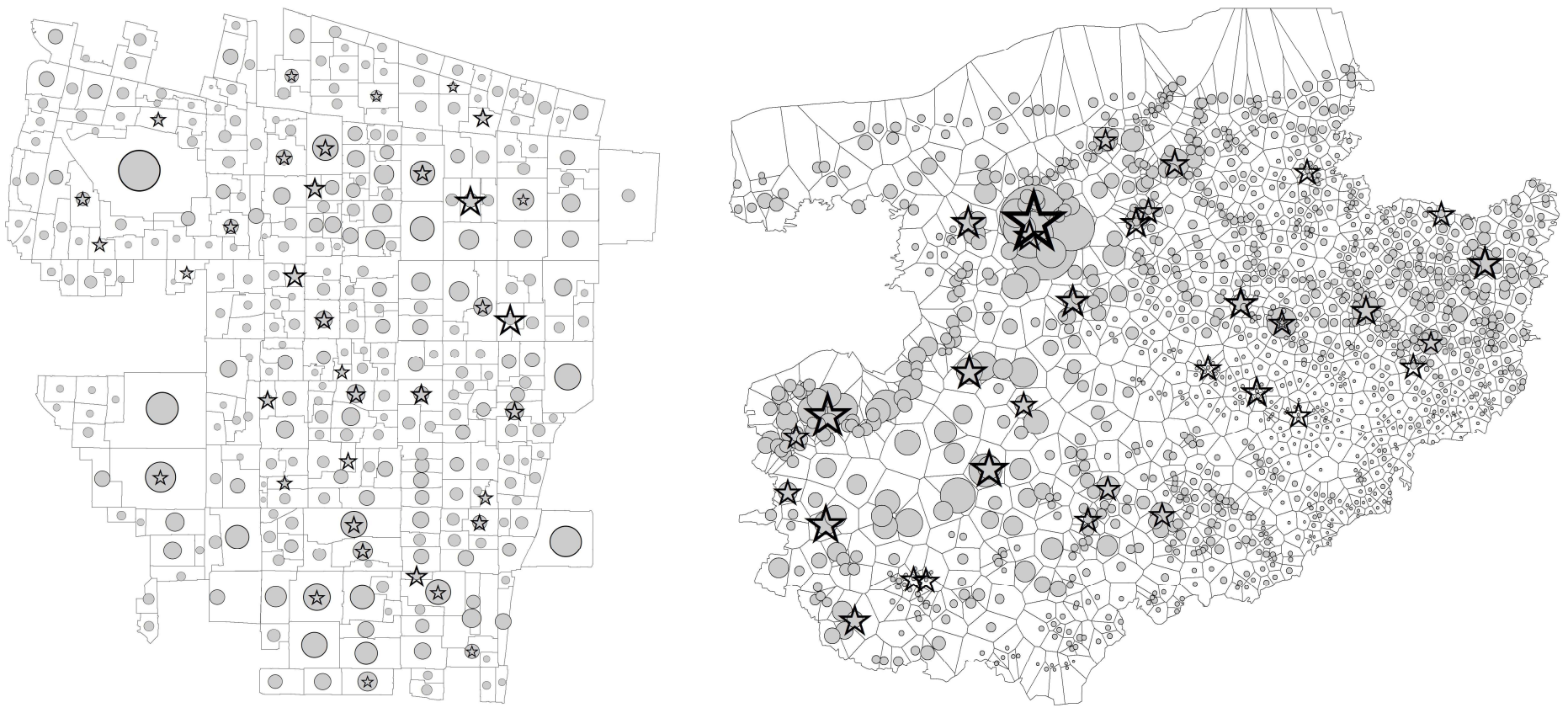

**Figure 1 Study areas ZY (left) and GY (right)**

Two sets of instances were prepared by changing the maximum facility capacities and facility fixed costs in the basic instances. For each instance, the maximum facility capacities were expanded by 20% and 40%, and the fixed facility costs were increased by 10%, 20%, 30% and 40%. As a results, in each study area, 15 instances can be created by combining different facility capacities and fixed costs. The dataset of new instances can be downloaded from webpage https://github.com/yfkong/Unified. The attributes of new instances are listed in Table 4. The column SDR illustrates the supply-demand ratio by dividing the total supply by the total demand, $SDR = \sum_{i \in I} s_i \,/\, \sum_{i \in j} d_j$. A tendency is that the smaller the ratio value, the more computation time is required for solving the instance (Gadegaard et al. 2018). The column CCR shows the ratio of total fixed facility cost to total of maximum capacities, $CCR = \sum_{i \in I} f_i \,/\, \sum_{i \in I} s_i$. Different fixed costs in objective function (1) have different effects on the selection of facility locations.

**Table 4 New instances for SSCFLP and its variants**

| Inst. | Group | $\|I\|$ | $\|J\|$ | SDR | CCR | Inst. | Group | $\|I\|$ | $\|J\|$ | SDR | CCR |
|---|---|---|---|---|---|---|---|---|---|---|---|
| ZYA1 | ZYA | 36 | 326 | 2.37 | 0.80 | GYA1 | GYA | 33 | 1276 | 1.61 | 1.83 |
| ZYA2 | ZYA | 36 | 326 | 2.37 | 0.88 | GYA2 | GYA | 33 | 1276 | 1.61 | 2.02 |
| ZYA3 | ZYA | 36 | 326 | 2.37 | 0.96 | GYA3 | GYA | 33 | 1276 | 1.61 | 2.20 |
| ZYA4 | ZYA | 36 | 326 | 2.37 | 1.04 | GYA4 | GYA | 33 | 1276 | 1.61 | 2.38 |
| ZYA5 | ZYA | 36 | 326 | 2.37 | 1.12 | GYA5 | GYA | 33 | 1276 | 1.61 | 2.56 |
| ZYB1 | ZYB | 36 | 326 | 2.84 | 0.67 | GYB1 | GYB | 33 | 1276 | 1.93 | 1.53 |
| ZYB2 | ZYB | 36 | 326 | 2.84 | 0.74 | GYB2 | GYB | 33 | 1276 | 1.93 | 1.68 |
| ZYB3 | ZYB | 36 | 326 | 2.84 | 0.80 | GYB3 | GYB | 33 | 1276 | 1.93 | 1.82 |
| ZYB4 | ZYB | 36 | 326 | 2.84 | 0.87 | GYB4 | GYB | 33 | 1276 | 1.93 | 1.98 |
| ZYB5 | ZYB | 36 | 326 | 2.84 | 0.94 | GYB5 | GYB | 33 | 1276 | 1.93 | 2.14 |
| ZYC1 | ZYC | 36 | 326 | 3.32 | 0.57 | GYC1 | GYC | 33 | 1276 | 2.26 | 1.31 |
| ZYC2 | ZYC | 36 | 326 | 3.32 | 0.63 | GYC2 | GYC | 33 | 1276 | 2.26 | 1.44 |
| ZYC3 | ZYC | 36 | 326 | 3.32 | 0.69 | GYC3 | GYC | 33 | 1276 | 2.26 | 1.57 |
| ZYC4 | ZYC | 36 | 326 | 3.32 | 0.75 | GYC4 | GYC | 33 | 1276 | 2.26 | 1.70 |
| ZYC5 | ZYC | 36 | 326 | 3.32 | 0.80 | GYC5 | GYC | 33 | 1276 | 2.26 | 1.83 |

The matheuristic algorithm's performance on SSCFLP and its variant problems is tested using the new instances. For SSCKFLP and CKFLSAP instances, the number of facilities *K* was set to 13~22 for ZY instances, and 16-25 for GY instances. Each instance was repeatedly solved for five times. The algorithm parameter *mloops* was set as 50 for all instances. To verify the optimality of the solutions, each instance was also solved by CPLEX Optimizer 12.6. The detailed solutions for new instances are shown in the appendix file of this article.

The exact and heuristic results for SSCFLP and it variants are summarized in Table 5. For each instance group, columns #opt, Gap and Time of CPLEX indicate the number of optimal solutions found by CPLEX, the average optimally gap (MIPGap), and the average computation time for each instance, respectively. The column Gap, Dev and Time of Matheuristic show the average optimal gap, the average relative standard deviation between the five solutions for each instance, and the average computation time for each instance.

There are several findings from Table 5. First, all the instances can be optimally or near-optimally solved by CPLEX. However, significant differences have been observed between different problem types, instance sizes, and supply-demand ratios. Since the constraints on spatial contiguity pose obstacles in solving the geographic problems, CFLSAP is harder to solve than SSCFLP in terms of solution optimality and computation time. On the other hand, SSCKFLP is easier to solve than SSCFLP, especially for ZY instances. Since a long computation time is required for most instances, it is a necessity to design heuristic methods for SSCFLP and its variants.

**Table 5 Summary of solutions from instances of SSCFLP and its variants**

| Problem | Inst. group | *K* | CPLEX | | | Matheuristic | | |
|---|---|---|---|---|---|---|---|---|
| | | | #opt | Gap/% | Time/s | Gap/% | Dev/% | Time/s |
| SSCFLP | ZYA | - | 5 | 0.00 | 1229.51 | 0.01 | 0.01 | 69.77 |
| SSCFLP | ZYB | - | 5 | 0.00 | 4367.56 | 0.37 | 0.13 | 75.21 |
| SSCFLP | ZYC | - | 5 | 0.00 | 394.97 | 0.02 | 0.04 | 89.35 |
| SSCFLP | GYA | - | 0 | 0.01 | 7200.00 | 0.01 | 0.01 | 271.20 |
| SSCFLP | GYB | - | 1 | 0.00 | 6296.33 | 0.00 | 0.00 | 241.19 |
| SSCFLP | GYC | - | 5 | 0.00 | 1588.71 | 0.00 | 0.01 | 235.64 |
| CFLSAP | ZYA | - | 0 | 0.26 | 7200.00 | 0.33 | 0.04 | 77.38 |
| CFLSAP | ZYB | - | 1 | 0.36 | 6366.91 | 0.38 | 0.03 | 92.15 |
| CFLSAP | ZYC | - | 2 | 0.04 | 6857.48 | 0.13 | 0.06 | 66.94 |
| CFLSAP | GYA | - | 0 | 0.19 | 7200.00 | 0.20 | 0.08 | 172.05 |
| CFLSAP | GYB | - | 2 | 0.13 | 6599.06 | 0.04 | 0.00 | 256.69 |
| CFLSAP | GYC | - | 0 | 0.22 | 7200.00 | 0.21 | 0.03 | 238.94 |
| SSCKFLP | ZYA1 | 13-22 | 10 | 0.00 | 161.32 | 0.10 | 0.08 | 59.55 |
| SSCKFLP | ZYB1 | 13-22 | 10 | 0.00 | 150.77 | 0.09 | 0.03 | 47.11 |
| SSCKFLP | ZYC1 | 13-22 | 10 | 0.00 | 27.66 | 0.01 | 0.03 | 49.42 |
| SSCKFLP | GYA1 | 16-25 | 5 | 0.01 | 3650.42 | 0.02 | 0.02 | 191.42 |
| SSCKFLP | GYB1 | 16-25 | 6 | 0.00 | 3669.45 | 0.01 | 0.01 | 194.36 |
| SSCKFLP | GYC1 | 16-25 | 9 | 0.00 | 1138.81 | 0.04 | 0.04 | 167.11 |
| CKFLSAP | ZYA1 | 13-22 | 8 | 0.01 | 4187.54 | 0.29 | 0.12 | 53.94 |
| CKFLSAP | ZYB1 | 13-22 | 10 | 0.00 | 2418.28 | 0.18 | 0.07 | 42.66 |
| CKFLSAP | ZYC1 | 13-22 | 10 | 0.00 | 453.68 | 0.08 | 0.07 | 35.48 |
| CKFLSAP | GYA1 | 16-25 | 4 | 0.07 | 5547.84 | 0.06 | 0.02 | 120.54 |
| CKFLSAP | GYB1 | 16-25 | 4 | 0.09 | 5969.06 | 0.10 | 0.02 | 145.04 |
| CKFLSAP | GYC1 | 16-25 | 6 | 0.06 | 4558.31 | 0.10 | 0.05 | 139.66 |

Second, the matheuristic algorithm is effective and efficient to solve SSCFLP and its variants with near-optimal gaps and small repeat deviations in a relatively short period of time. The solutions found by the proposed algorithm approximate optimal solutions or the lower bounds with average gaps of 0.07% for SSCFLP, 0.22% for CFLSAP, 0.04% for SSCKFLP, and 0.13% for CKFLSAP. For many CFLSAP GY instances, the solutions of the matheuristic in several minutes of computation time are better than those of CPLEX in two hours of computation time. The relative deviation of the solutions from repeatedly executing the matheuristic algorithm is rather small, ranging from 0.00% to 0.13%. It is also found that the constraints on spatial contiguity has much effects on the computation time of CPLEX, but has little effect on that of the matheuristic algorithm.

Comparison of SSCFLP solutions and CFLSAP solutions on the same instances show that three are substantial differences between them. The objectives of CFLSAP instances increase slightly, by a range of between 0.09% and 1.00% for ZY instances, and a range of between 0.01% and 0.36% for GY instances. However, some service areas in SSCFLP solutions are not contiguous; but all the service areas in CFLSAP solutions are guaranteed to be contiguous. More importantly, the facility locations in SSCFLP solutions may be very different from that in CFLSAP solutions. Figure 2 show the best known SSCFLP solution and CFLSAP solution from instance ZYA4. The objectives are 5327.60 and 5344.66 for SSCFLP and CFLSAP, respectively. In SSCFLP solution, three service areas are not contiguous. More importantly, the facility locations and their service areas in SSCFLP solution are largely different from those in CFLSAP solution.

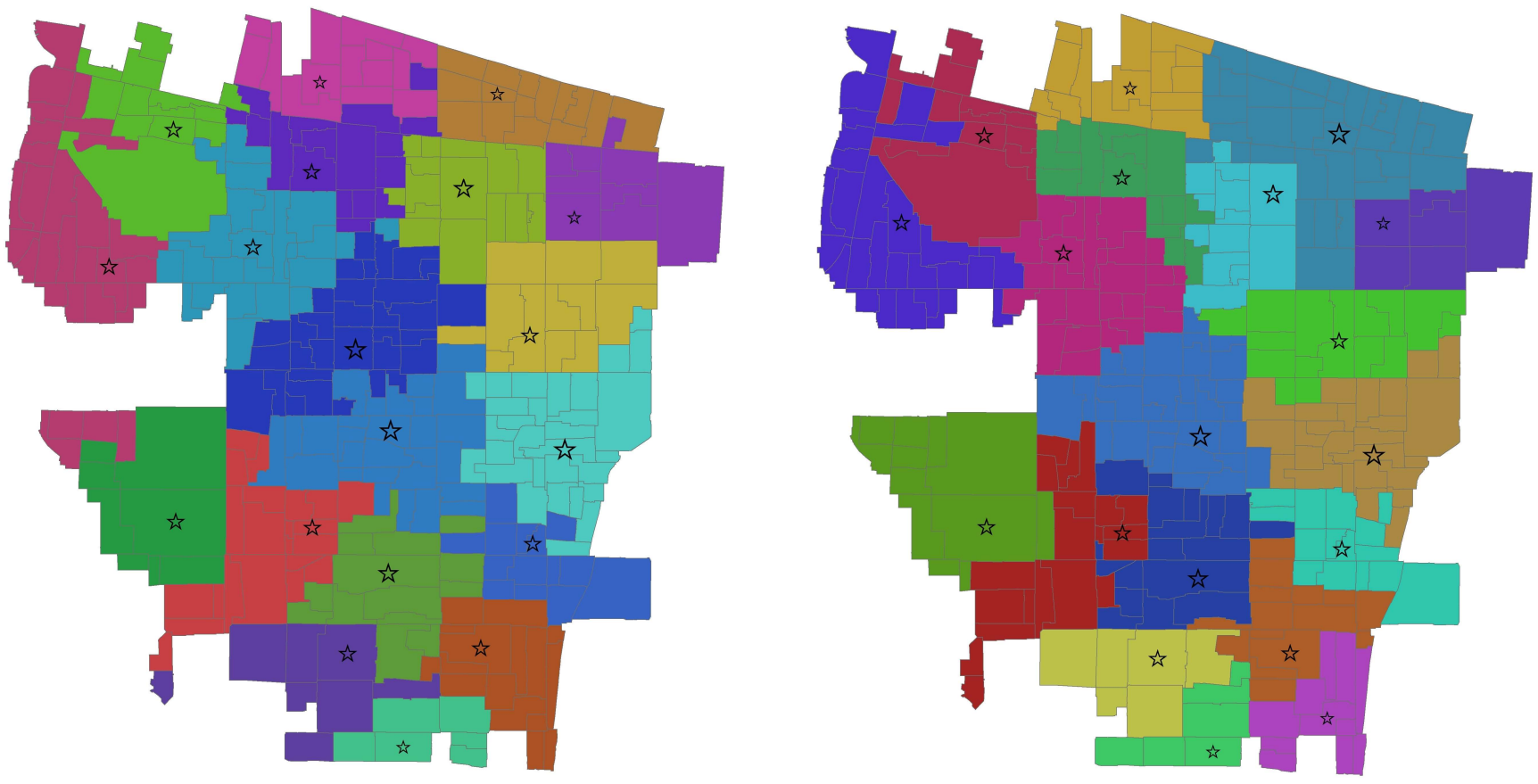

**Figure 2 The best-known SSCFLP solution (left) and CFLSAP solution (right) from instance ZYA4**

SSCKFLP and CKFSAP solutions show that the constraint of the number of facilities has great effects on the location selection and the cost objective. For a specific SSCFLP instance, setting a constraint with more or less facilities than the optimal number of facility locations will not only change the selection of locations, but also the cost objective. Figure 3 shows the variation of objectives with the number of facilities for instance ZYA1 and instance GYC1. For SSCFLP, the optimal numbers of facilities are 19 and 20 for the two instances, respectively. It is observed that, for instance ZYA1, the solution objectives increase by 0.17%~20.43% when the parameter $K$ is set to a number other than 19. Similarly, for instance GYC1, the solution objectives increase by 0.06%~4.33% when $K$ is set to a number other than 20.

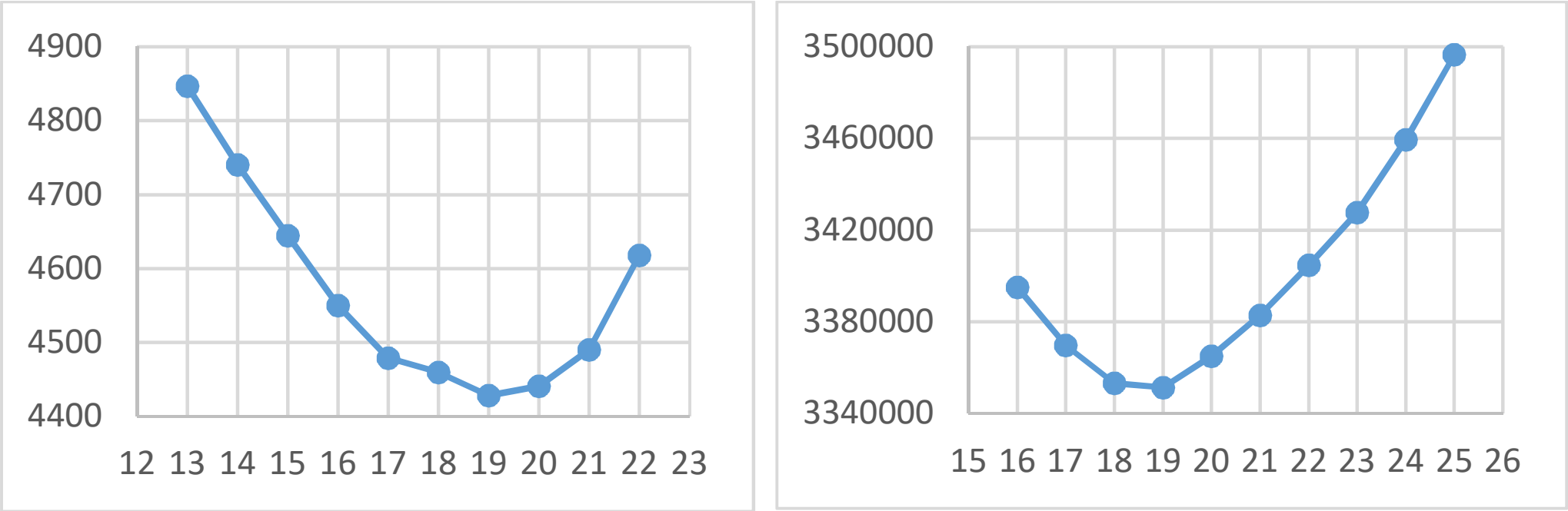

**Figure 3 SSCFKLP objectives (vertical axis) versus the number of facilities (horizontal axis): instance ZYA1 (left) and instance GYC1 (right)**

# 5 Conclusions

In this article, three SSCFLP variants are defined by extending SSCFLP with the contiguity constraints on facility service areas and/or the constraint of the number of facilities. CFLSAP and CKFLSAP are formulated as mixed integer linear programs by embedding a network flow-based model into the classical SSCFLP model. CFLSAP and CKFLSAP instances with 33 candidate locations and 1276 customers can be optimally or near-optimally solved by CPLEX.

A matheuristic algorithm is proposed for the single-source capacitated facility location problem (SSCFLP) and its variants. It starts from an initial solution, and iteratively improves the solution by mathematically solving large neighborhood-based sub-problems. The performance of the algorithm is tested on 5 well-known sets of SSCFLP benchmark instances. Among the 272 instances, 191 optimal solutions are found, and 35 best-known solutions are updated. For the largest set of instances in Avella and Boccia (2009), the proposed algorithm outperforms the state-of-the-art methods in terms of the solution quality and the computational time. Furthermore, based on two geographic areas, two sets of instances are generated to test the algorithm for solving SSCFLP and its variants. The solutions found by the proposed algorithm approximate optimal solutions or the lower bounds with average gaps of 0.07% for SSCFLP, 0.22% for CFLSAP, 0.04% for SSCKFLP, and 0.13% for CKFLSAP.

These are three findings from the solution results of SSCFLP and its variants. First, SSCFLP is NP-Hard in strong sense, and thus is hard to solve by exact methods. Adding new constraints on spatial contiguity in SSCFLP, CFLSAP is much more difficult to solve. However, adding a constraint on the number of facilities, SSCKFLP is relatively easy to solve for most instances. Second, the matheuristic algorithm can be used to solve SSCFLP and its variants effectively and efficiently in a computation time of several minutes. Its performance is not substantively effected by the additional constraints. Third, the facility locations, service areas, and cost objective for SSCFLP and it variants are sensitive to instance features such as the supply-demand ratio, the facility cost-capacity ratio, the contiguity constraints on service areas and the constraint of the number of facilities.

Open issues still remain in this research. The proposed matheuristic algorithm outperforms the state-of-the-art algorithms for solving medium-size and large-size instances. However, it is not the best algorithm for solving small-size OR-Lib and Holmberg instances. Experimentation shows that the small-size SSCFLP instances is easy to solve by CPLEX branch-and-cut method directly. Therefore, it is not necessary to solve it by repeatedly exploring its neighborhood. The future research should investigate the computational complexity of an instance and thus decide which method, matheuristic or branch-and-cut, is the best choice to solve the instance. On the other hand, for large-size instances, the performance of the matheuristic depends on the size of neighborhood. It is efficient to solve the small neighborhood-based sub-problems, however, the current solution may be improved with a low possibility. The solution can be improved easily by solving the sub-problems related to very large neighborhood, however, such sub-problem models is not easy to solve. It is essential to investigate the best choice of parameters $Q$ and $U_{max}$ in the matheuristic algorithm.

## Acknowledgments

Research partially supported by the National Natural Science Foundation of China (No. 41871307).

# Appendix: Computational results

Note:

1 All the computational results in this article were obtained from a HP desktop computer with Intel Core I7-6700 CPU 3.40 GHz, 8 GB RAM and the Windows 10 operating system.

2 In all tables, the optimal values are highlighted in red.

3 In all tables, columns are explained as follows:

- LB: lower bound obtained by CPLEX;
- UB: upper bound obtained by CPLEX;
- Gap: the gap between solution objective and the lower bound;
- Time: solution time in seconds.
- Objmin: minimum objective;
- Objavg: average objective;
- Gapavg: the gap between average objective and lower bound;
- Stdev: relative standard deviation of the five objectives from an instance.

**Table 1: Solution results from Yang Instances**

| Dataset | Instance | Optimal | Objmin | Objavg | Gapavg | Stdev | Time |
|---|---|---|---|---|---|---|---|
| Yang | 30_200_1 | 30181 | 30181 | 30181.6 | 0.00% | 0.00% | 141.764 |
| Yang | 30_200_2 | 28923 | 28923 | 28926.4 | 0.01% | 0.03% | 101.445 |
| Yang | 30_200_3 | 28131 | 28131 | 28131.0 | 0.00% | 0.00% | 35.897 |
| Yang | 30_200_4 | 28152 | 28152 | 28152.0 | 0.00% | 0.00% | 91.1632 |
| Yang | 30_200_5 | 27646 | 27646 | 27646.0 | 0.00% | 0.00% | 11.2858 |
| Yang | 60_200_1 | 27977 | 27977 | 27977.0 | 0.00% | 0.00% | 151.518 |
| Yang | 60_200_2 | 29704 | 29709 | 29709.0 | 0.02% | 0.00% | 142.003 |
| Yang | 60_200_3 | 27993 | 27993 | 27993.0 | 0.00% | 0.00% | 99.3634 |
| Yang | 60_200_4 | 27691 | 27691 | 27691.4 | 0.00% | 0.00% | 125.465 |
| Yang | 60_200_5 | 29195 | 29195 | 29205.0 | 0.03% | 0.05% | 125.571 |
| Yang | 60_300_1 | 35648 | 35650 | 35692.2 | 0.12% | 0.08% | 190.705 |
| Yang | 60_300_2 | 35474 | 35474 | 35474.6 | 0.00% | 0.00% | 53.004 |
| Yang | 60_300_3 | 33872 | 33872 | 33872.0 | 0.00% | 0.00% | 104.625 |
| Yang | 60_300_4 | 33096 | 33096 | 33096.0 | 0.00% | 0.00% | 129.968 |
| Yang | 60_300_5 | 30918 | 30918 | 30947.2 | 0.09% | 0.06% | 124.594 |
| Yang | 80_400_1 | 39318 | 39318 | 39432.4 | 0.29% | 0.39% | 334.89 |
| Yang | 80_400_2 | 37076 | 37076 | 37076.0 | 0.00% | 0.00% | 154.372 |
| Yang | 80_400_3 | 43859 | 43918 | 43923.2 | 0.15% | 0.02% | 318.489 |
| Yang | 80_400_4 | 37344 | 37344 | 37344.0 | 0.00% | 0.00% | 91.9016 |
| Yang | 80_400_5 | 43508 | 43508 | 43510.4 | 0.01% | 0.01% | 264.298 |

### Table 2: Solution results from TB4 Instances

| Inst. | Opt/LB | Objmin | Objavg | Gapavg | Stdev | Time/s |
|---|---|---|---|---|---|---|
| 50_100_2_1 | 18294.00 | 18294.00 | 18294.80 | 0.00% | 0.01% | 38.43 |
| 50_100_2_2 | 19688.00 | 19690.00 | 19705.00 | 0.09% | 0.04% | 54.89 |
| 50_100_2_3 | 19075.00 | 19075.00 | 19102.00 | 0.14% | 0.09% | 65.34 |
| 50_100_2_4 | 18620.00 | 18620.00 | 18624.40 | 0.02% | 0.04% | 60.32 |
| 50_100_2_5 | 18502.00 | 18517.00 | 18523.20 | 0.11% | 0.02% | 145.44 |
| 50_100_3_1 | 16948.00 | 16949.00 | 16952.00 | 0.02% | 0.02% | 242.69 |
| 50_100_3_2 | 15063.00 | 15063.00 | 15063.40 | 0.00% | 0.00% | 61.63 |
| 50_100_3_3 | 15107.00 | 15107.00 | 15107.00 | 0.00% | 0.00% | 25.95 |
| 50_100_3_4 | 14347.00 | 14347.00 | 14347.00 | 0.00% | 0.00% | 114.43 |
| 50_100_3_5 | 14813.00 | 14813.00 | 14817.80 | 0.03% | 0.03% | 242.11 |
| 50_100_5_1 | 12072.00 | 12072.00 | 12072.00 | 0.00% | 0.00% | 26.41 |
| 50_100_5_2 | 11898.00 | 11898.00 | 11902.80 | 0.04% | 0.02% | 113.31 |
| 50_100_5_3 | 11125.00 | 11125.00 | 11125.00 | 0.00% | 0.00% | 54.03 |
| 50_100_5_4 | 11817.00 | 11817.00 | 11817.00 | 0.00% | 0.00% | 43.54 |
| 50_100_5_5 | 11489.00 | 11489.00 | 11489.00 | 0.00% | 0.00% | 9.70 |
| 50_200_2_1 | 25992.00 | 25995.00 | 25995.60 | 0.01% | 0.00% | 38.53 |
| 50_200_2_2 | 25868.00 | 25868.00 | 25868.60 | 0.00% | 0.00% | 41.73 |
| 50_200_2_3 | 26930.00 | 26931.00 | 26950.40 | 0.08% | 0.04% | 94.25 |
| 50_200_2_4 | 25951.63 | 25954.00 | 25954.60 | 0.01% | 0.00% | 33.25 |
| 50_200_2_5 | 25326.50 | 25329.00 | 25378.80 | 0.21% | 0.11% | 53.61 |
| 50_200_3_1 | 20701.00 | 20702.00 | 20702.00 | 0.00% | 0.00% | 60.29 |
| 50_200_3_2 | 22021.00 | 22021.00 | 22052.40 | 0.14% | 0.13% | 125.42 |
| 50_200_3_3 | 20038.00 | 20038.00 | 20038.00 | 0.00% | 0.00% | 16.69 |
| 50_200_3_4 | 20595.00 | 20596.00 | 20596.00 | 0.00% | 0.00% | 17.36 |
| 50_200_3_5 | 21168.00 | 21168.00 | 21168.20 | 0.00% | 0.00% | 29.36 |
| 50_200_5_1 | 16659.00 | 16659.00 | 16765.40 | 0.64% | 1.41% | 49.14 |
| 50_200_5_2 | 16138.00 | 16138.00 | 16138.00 | 0.00% | 0.00% | 42.64 |
| 50_200_5_3 | 17755.00 | 17755.00 | 17755.00 | 0.00% | 0.00% | 56.29 |
| 50_200_5_4 | 15858.00 | 15858.00 | 15858.00 | 0.00% | 0.00% | 49.91 |
| 50_200_5_5 | 16884.00 | 16885.00 | 16885.00 | 0.01% | 0.00% | 8.98 |
| 60_300_2_1 | 34,858.50 | 34861.00 | 34862.40 | 0.01% | 0.00% | 56.90 |
| 60_300_2_2 | 36,543.50 | 36552.00 | 36570.40 | 0.07% | 0.08% | 81.05 |
| 60_300_2_3 | 34,876.20 | 34879.00 | 34879.80 | 0.01% | 0.00% | 64.60 |
| 60_300_2_4 | 34,817.60 | 34823.00 | 34863.40 | 0.13% | 0.06% | 52.48 |
| 60_300_2_5 | 37,138.10 | 37142.00 | 37156.40 | 0.05% | 0.06% | 68.20 |
| 60_300_3_1 | 27903.00 | 27904.00 | 27904.80 | 0.01% | 0.00% | 31.62 |
| 60_300_3_2 | 27594.00 | 27594.00 | 27596.00 | 0.01% | 0.01% | 36.00 |
| 60_300_3_3 | 29231.00 | 29232.00 | 29382.20 | 0.52% | 1.09% | 107.33 |
| 60_300_3_4 | 27439.00 | 27439.00 | 27439.00 | 0.00% | 0.00% | 25.28 |
| 60_300_3_5 | 28033.00 | 28052.00 | 28078.40 | 0.16% | 0.06% | 78.34 |
| 60_300_5_1 | 21045.00 | 21045.00 | 21045.00 | 0.00% | 0.00% | 55.91 |
| 60_300_5_2 | 22589.00 | 22589.00 | 22589.00 | 0.00% | 0.00% | 112.26 |
| 60_300_5_3 | 21449.00 | 21449.00 | 21463.00 | 0.07% | 0.15% | 76.93 |
| 60_300_5_4 | 21466.00 | 21466.00 | 21466.00 | 0.00% | 0.00% | 18.35 |
| 60_300_5_5 | 21860.00 | 21860.00 | 21860.00 | 0.00% | 0.00% | 58.60 |

## Table 3: Solution results from Tbed1 Instances

| Instance | LB | NewLB | Objmin | Objavg | Gapavg | Stdev | Time |
|---|---|---|---|---|---|---|---|
| i300_1 | 16552.75 | 16555.77 | 16563.14 | 16572.00 | 0.10% | 0.04% | 71.74 |
| i300_2 | 16059.34 | 16059.34 | 16135.82 | 16156.56 | 0.61% | 0.14% | 91.99 |
| i300_3 | 15606.10 | 15606.10 | 15666.23 | 15681.61 | 0.48% | 0.13% | 69.21 |
| i300_4 | 18143.89 | 18143.89 | 18255.10 | 18276.83 | 0.73% | 0.09% | 82.86 |
| i300_5 | 18191.11 | 18191.11 | 18291.05 | 18305.47 | 0.63% | 0.12% | 142.50 |
| i300_6 | 11271.22 | 11326.43 | 11329.93 | 11338.37 | 0.11% | 0.08% | 55.25 |
| i300_7 | 11461.16 | 11470.31 | 11473.11 | 11477.61 | 0.06% | 0.05% | 53.33 |
| i300_8 | 11449.67 | 11449.67 | 11455.19 | 11457.13 | 0.07% | 0.04% | 48.17 |
| i300_9 | 10932.88 | 10932.88 | 10932.88 | 10933.45 | 0.01% | 0.01% | 60.32 |
| i300_10 | 11324.34 | 11324.34 | 11324.34 | 11327.35 | 0.03% | 0.04% | 46.39 |
| i300_11 | 10046.94 | 10046.94 | 10050.50 | 10050.88 | 0.04% | 0.01% | 46.28 |
| i300_12 | 9359.64 | 9359.64 | 9359.64 | 9359.64 | 0.00% | 0.00% | 38.93 |
| i300_13 | 10103.49 | 10103.49 | 10103.49 | 10107.94 | 0.04% | 0.07% | 59.62 |
| i300_14 | 9738.05 | 9738.05 | 9738.05 | 9742.70 | 0.05% | 0.07% | 48.04 |
| i300_15 | 9902.26 | 9902.26 | 9902.26 | 9902.71 | 0.00% | 0.01% | 48.42 |
| i300_16 | 9168.08 | 9168.08 | 9168.08 | 9169.53 | 0.02% | 0.01% | 44.52 |
| i300_17 | 9181.07 | 9181.07 | 9181.07 | 9181.07 | 0.00% | 0.00% | 39.78 |
| i300_18 | 9581.95 | 9581.95 | 9581.95 | 9586.16 | 0.04% | 0.06% | 56.14 |
| i300_19 | 9062.16 | 9062.16 | 9062.16 | 9064.14 | 0.02% | 0.05% | 39.79 |
| i300_20 | 9077.85 | 9077.85 | 9077.85 | 9079.80 | 0.02% | 0.05% | 38.63 |
| i3001500_1 | 154999.14 | 154999.14 | 154999.19 | 155009.37 | 0.01% | 0.00% | 96.48 |
| i3001500_2 | 159438.03 | 159438.03 | 159446.39 | 159451.81 | 0.01% | 0.01% | 104.29 |
| i3001500_3 | 157300.15 | 157300.15 | 157306.78 | 157313.99 | 0.01% | 0.00% | 98.95 |
| i3001500_4 | 157796.28 | 157796.28 | 157797.17 | 157799.99 | 0.00% | 0.00% | 101.32 |
| i3001500_5 | 161305.97 | 161305.97 | 161305.97 | 161326.82 | 0.01% | 0.01% | 109.75 |
| i3001500_6 | 156667.31 | 156667.31 | 156667.31 | 156667.31 | 0.00% | 0.00% | 111.09 |
| i3001500_7 | 157031.55 | 157031.55 | 157031.55 | 157031.98 | 0.00% | 0.00% | 81.20 |
| i3001500_8 | 157796.21 | 157796.21 | 157796.21 | 157800.09 | 0.00% | 0.00% | 96.64 |
| i3001500_9 | 156968.46 | 156968.46 | 156968.46 | 156968.46 | 0.00% | 0.00% | 96.49 |
| i3001500_10 | 157757.62 | 157757.62 | 157757.62 | 157758.98 | 0.00% | 0.00% | 92.57 |
| i3001500_11 | 150015.13 | 150015.13 | 150015.13 | 150015.62 | 0.00% | 0.00% | 68.05 |
| i3001500_12 | 154937.67 | 154937.67 | 154937.67 | 154939.25 | 0.00% | 0.00% | 87.14 |
| i3001500_13 | 151608.42 | 151608.42 | 151608.42 | 151610.47 | 0.00% | 0.00% | 85.19 |
| i3001500_14 | 151848.05 | 151848.05 | 151848.05 | 151848.05 | 0.00% | 0.00% | 71.85 |
| i3001500_15 | 156480.89 | 156480.89 | 156480.89 | 156483.93 | 0.00% | 0.00% | 84.68 |
| i3001500_16 | 155493.77 | 155493.77 | 155493.77 | 155493.77 | 0.00% | 0.00% | 70.04 |
| i3001500_17 | 156038.04 | 156038.04 | 156038.04 | 156041.23 | 0.00% | 0.00% | 71.02 |
| i3001500_18 | 156790.75 | 156790.75 | 156790.75 | 156790.88 | 0.00% | 0.00% | 75.33 |
| i3001500_19 | 155947.13 | 155947.13 | 155947.13 | 155947.13 | 0.00% | 0.00% | 74.52 |
| i3001500_20 | 156426.14 | 156426.14 | 156426.14 | 156427.78 | 0.00% | 0.00% | 75.10 |
| i500_1 | 26566.69 | 26566.69 | 26731.63 | 26755.86 | 0.71% | 0.09% | 154.35 |
| i500_2 | 28268.41 | 28268.41 | 28460.43 | 28503.23 | 0.83% | 0.12% | 172.13 |
| i500_3 | 28067.69 | 28067.69 | 28284.36 | 28296.54 | 0.82% | 0.03% | 157.79 |
| i500_4 | 28268.36 | 28268.36 | 28489.69 | 28506.64 | 0.84% | 0.11% | 216.23 |
| i500_5 | 24805.56 | 24805.56 | 24995.73 | 25010.40 | 0.83% | 0.08% | 138.33 |
| i500_6 | 15842.27 | 15853.35 | 15867.59 | 15872.42 | 0.12% | 0.03% | 112.62 |
| i500_7 | 16163.11 | 16205.15 | 16205.15 | 16239.53 | 0.21% | 0.20% | 119.70 |
| i500_8 | 16081.54 | 16081.54 | 16125.30 | 16126.99 | 0.28% | 0.01% | 139.58 |
| i500_9 | 16346.25 | 16399.40 | 16430.98 | 16431.07 | 0.19% | 0.00% | 136.64 |
| i500_10 | 15857.97 | 15857.97 | 15887.80 | 15888.47 | 0.19% | 0.01% | 115.30 |
| i500_11 | 13497.71 | 13497.71 | 13504.94 | 13505.64 | 0.06% | 0.00% | 121.35 |
| i500_12 | 14736.38 | 14736.38 | 14736.38 | 14761.07 | 0.17% | 0.19% | 189.63 |
| i500_13 | 13709.76 | 13715.96 | 13715.96 | 13719.07 | 0.02% | 0.02% | 142.29 |
| i500_14 | 13629.54 | 13629.54 | 13634.91 | 13643.75 | 0.10% | 0.07% | 148.73 |
| i500_15 | 13940.08 | 13947.12 | 13947.38 | 13947.59 | 0.00% | 0.00% | 154.86 |

**Table 3: Solution results from Tbed1 Instances (continued)**

| Instance | LB | NewLB | Objmin | Objavg | Gapavg | Stdev | Time |
|---|---|---|---|---|---|---|---|
| i500_16 | 12618.68 | 12618.68 | 12618.68 | 12622.25 | 0.03% | 0.03% | 196.72 |
| i500_17 | 13386.17 | 13386.17 | 13386.17 | 13388.37 | 0.02% | 0.02% | 205.13 |
| i500_18 | 12852.52 | 12852.52 | 12852.52 | 12852.52 | 0.00% | 0.00% | 194.16 |
| i500_19 | 13521.52 | 13521.52 | 13521.59 | 13521.59 | 0.00% | 0.00% | 160.83 |
| i500_20 | 12362.20 | 12362.20 | 12362.20 | 12362.20 | 0.00% | 0.00% | 197.48 |
| i700_1 | 37054.60 | 37054.60 | 37343.34 | 37351.36 | 0.80% | 0.02% | 187.87 |
| i700_2 | 34488.56 | 34488.56 | 34817.44 | 34824.04 | 0.97% | 0.03% | 258.69 |
| i700_3 | 34485.24 | 34485.24 | 34759.07 | 34796.56 | 0.90% | 0.13% | 270.84 |
| i700_4 | 38260.98 | 38260.98 | 38534.35 | 38595.24 | 0.87% | 0.09% | 221.20 |
| i700_5 | 37950.49 | 37950.49 | 38230.13 | 38280.86 | 0.87% | 0.08% | 229.96 |
| i700_6 | 19881.11 | 19910.67 | 20065.19 | 20071.69 | 0.81% | 0.04% | 265.27 |
| i700_7 | 21295.25 | 21297.30 | 21433.56 | 21445.29 | 0.69% | 0.08% | 266.35 |
| i700_8 | 20702.95 | 20702.95 | 20820.70 | 20829.25 | 0.61% | 0.07% | 198.95 |
| i700_9 | 20976.65 | 20979.88 | 21104.52 | 21106.78 | 0.60% | 0.02% | 224.51 |
| i700_10 | 22039.40 | 22055.41 | 22210.50 | 22212.42 | 0.71% | 0.01% | 231.37 |
| i700_11 | 17105.77 | 17120.15 | 17188.47 | 17194.92 | 0.44% | 0.04% | 644.55 |
| i700_12 | 18135.97 | 18135.97 | 18201.06 | 18206.63 | 0.39% | 0.07% | 468.79 |
| i700_13 | 17277.92 | 17277.92 | 17299.20 | 17316.70 | 0.22% | 0.22% | 417.62 |
| i700_14 | 17374.91 | 17383.87 | 17383.87 | 17383.87 | 0.00% | 0.00% | 332.48 |
| i700_15 | 18167.98 | 18167.98 | 18220.75 | 18223.55 | 0.31% | 0.02% | 455.16 |
| i700_16 | 16029.55 | 16029.55 | 16029.55 | 16029.55 | 0.00% | 0.00% | 408.28 |
| i700_17 | 16199.55 | 16199.55 | 16199.55 | 16206.53 | 0.04% | 0.02% | 387.65 |
| i700_18 | 16443.07 | 16443.07 | 16443.07 | 16443.54 | 0.00% | 0.00% | 419.71 |
| i700_19 | 16399.79 | 16399.79 | 16407.94 | 16415.00 | 0.09% | 0.04% | 451.81 |
| i700_20 | 15476.99 | 15492.02 | 15492.02 | 15494.65 | 0.02% | 0.02% | 482.59 |
| i1000_1 | 49681.02 | 49681.02 | 50104.98 | 50125.83 | 0.90% | 0.04% | 389.31 |
| i1000_2 | 50842.16 | 50842.16 | 51277.80 | 51322.54 | 0.94% | 0.11% | 389.47 |
| i1000_3 | 47362.62 | 47362.62 | 47737.73 | 47788.72 | 0.90% | 0.07% | 460.01 |
| i1000_4 | 49029.12 | 49029.12 | 49408.86 | 49483.82 | 0.93% | 0.15% | 472.67 |
| i1000_5 | 50971.44 | 50971.44 | 51415.41 | 51449.69 | 0.94% | 0.07% | 367.96 |
| i1000_6 | 27804.13 | 27823.84 | 28043.73 | 28068.16 | 0.88% | 0.08% | 241.85 |
| i1000_7 | 27210.23 | 27252.32 | 27412.21 | 27450.79 | 0.73% | 0.08% | 229.67 |
| i1000_8 | 27307.81 | 27375.37 | 27543.01 | 27560.39 | 0.68% | 0.09% | 279.79 |
| i1000_9 | 26816.77 | 26857.09 | 26992.81 | 27028.92 | 0.64% | 0.10% | 240.04 |
| i1000_10 | 27178.05 | 27186.99 | 27397.38 | 27413.84 | 0.83% | 0.04% | 237.51 |
| i1000_11 | 22117.85 | 22180.33 | 22247.61 | 22259.53 | 0.36% | 0.06% | 351.24 |
| i1000_12 | 22110.95 | 22160.39 | 22231.18 | 22236.76 | 0.34% | 0.04% | 363.93 |
| i1000_13 | 22592.90 | 22657.09 | 22745.61 | 22776.41 | 0.53% | 0.11% | 456.50 |
| i1000_14 | 22273.28 | 22312.01 | 22405.92 | 22435.77 | 0.55% | 0.08% | 289.20 |
| i1000_15 | 22572.17 | 22629.44 | 22704.67 | 22715.42 | 0.38% | 0.04% | 336.20 |
| i1000_16 | 21322.81 | 21331.81 | 21390.24 | 21390.89 | 0.28% | 0.01% | 331.09 |
| i1000_17 | 21209.83 | 21209.83 | 21234.62 | 21250.19 | 0.19% | 0.04% | 347.68 |
| i1000_18 | 20739.20 | 20739.20 | 20753.39 | 20767.32 | 0.14% | 0.05% | 376.15 |
| i1000_19 | 20529.02 | 20537.45 | 20597.01 | 20610.08 | 0.35% | 0.06% | 487.19 |
| i1000_20 | 21541.51 | 21560.86 | 21601.55 | 21603.30 | 0.20% | 0.01% | 264.22 |

**Table 4: Correction of solutions obtained by CPLEX 12.6**

| Dataset | Instance | LB | UB | Memo |
|---|---|---|---|---|
| TB4 | 50-200-2-4 | 25951.63 | 25955 | Correct Gadegaard et al. (2018) |
| TB4 | 50-200-2-5 | 25326.50 | 25329 | Correct Gadegaard et al. (2018) |
| TB4 | 50-300-2-5 | 37138.10 | 37142 | Correct Gadegaard et al. (2018) |

**Table 5: New optimal solutions obtained by CPLEX 12.6**

| Dataset | Instance | LB | UB | Memo |
|---|---|---|---|---|
| Tbed1 | i300_6 | 11326.43 | 11326.43 | Update Caserta & Voß (2020) |
| Tbed1 | i300_7 | 11470.31 | 11470.31 | Update Caserta & Voß (2020) |
| Tbed1 | i500_6 | 15853.35 | 15853.35 | Update Caserta & Voß (2020) |
| Tbed1 | i500_7 | 16205.15 | 16205.15 | Update Caserta & Voß (2020) |
| Tbed1 | i500_9 | 16399.40 | 16399.4 | Update Caserta & Voß (2020) |
| Tbed1 | i500_10 | 15886.54 | 15886.54 | Update Caserta & Voß (2020) |
| Tbed1 | i500_13 | 13715.96 | 13715.96 | Update Caserta & Voß (2020) |
| Tbed1 | i500_15 | 13947.12 | 13947.12 | Update Caserta & Voß (2020) |
| Tbed1 | i700_14 | 17383.87 | 17383.87 | Update Caserta & Voß (2020) |
| Tbed1 | i700_20 | 15492.02 | 15492.02 | Update Caserta & Voß (2020) |

## Table 6: New best known solutions

| Dataset | instance | LB | BestKnown | NewBestKnown | Improvement |
|---|---|---|---|---|---|
| TB4 | 60-300-2-1 | 34858.5 | 34861 | 34860 | 0.00% |
| TB4 | 60-300-2-2 | 36543.5 | 36742 | 36551 | 0.52% |
| TB4 | 60-300-2-3 | 34876.2 | 34884 | 34878 | 0.02% |
| TB4 | 60-300-2-4 | 34817.6 | 36057 | 34821 | 3.43% |
| Tbed1 | i300_2 | 16059.34 | 16140.*00* | 16135.82 | 0.03% |
| Tbed1 | i300_3 | 15606.10 | 15687.38 | 15666.23 | 0.13% |
| Tbed1 | i300_4 | 18143.89 | 18312.60 | 18255.10 | 0.31% |
| Tbed1 | i300_5 | 18191.11 | 18315.44 | 18291.05 | 0.13% |
| Tbed1 | i500_1 | 26566.69 | 26824.08 | 26731.63 | 0.34% |
| Tbed1 | i500_3 | 28067.68 | 28362.79 | 28284.36 | 0.28% |
| Tbed1 | i500_4 | 28268.36 | 28518.40 | 28489.69 | 0.10% |
| Tbed1 | i700_1 | 37054.60 | 37751.08 | 37343.34 | 1.08% |
| Tbed1 | i700_2 | 34488.56 | 35076.83 | 34817.44 | 0.74% |
| Tbed1 | i700_3 | 34485.24 | 34977.47 | 34759.07 | 0.62% |
| Tbed1 | i700_4 | 38260.98 | 38860.34 | 38534.35 | 0.84% |
| Tbed1 | i700_7 | 21297.30 | 21437.82 | 21433.56 | 0.02% |
| Tbed1 | i700_8 | 20659.96 | 20823.75 | 20820.70 | 0.01% |
| Tbed1 | i700_10 | 22055.41 | 22274.57 | 22210.50 | 0.29% |
| Tbed1 | i700_11 | 17120.15 | 17189.64 | 17188.47 | 0.01% |
| Tbed1 | i700_12 | 18135.97 | 18232.53 | 18201.06 | 0.17% |
| Tbed1 | i1000_1 | 49681.02 | 50734.33 | 50104.98 | 1.24% |
| Tbed1 | i1000_2 | 50842.16 | 51677.00 | 51277.80 | 0.77% |
| Tbed1 | i1000_3 | 47362.62 | 48141.82 | 47737.73 | 0.84% |
| Tbed1 | i1000_4 | 49029.12 | 49910.85 | 49408.86 | 1.01% |
| Tbed1 | i1000_5 | 50971.44 | 51824.38 | 51415.41 | 0.79% |
| Tbed1 | i1000_6 | 27823.84 | 28051.58 | 28043.73 | 0.03% |
| Tbed1 | i1000_7 | 27252.32 | 27521.50 | 27412.21 | 0.40% |
| Tbed1 | i1000_8 | 27375.37 | 27638.39 | 27543.01 | 0.35% |
| Tbed1 | i1000_9 | 26857.09 | 27127.70 | 26992.81 | 0.50% |
| Tbed1 | i1000_10 | 27186.99 | 27469.49 | 27399.38 | 0.26% |
| Tbed1 | i1000_11 | 22180.33 | 22297.32 | 22247.61 | 0.22% |
| Tbed1 | i1000_12 | 22160.39 | 22231.34 | 22231.18 | 0.00% |
| Tbed1 | i1000_13 | 22657.09 | 22768.69 | 22745.61 | 0.10% |
| Tbed1 | i1000_15 | 22629.44 | 22706.59 | 22704.67 | 0.01% |
| Tbed1 | i1000_20 | 21560.86 | 21618.06 | 21601.55 | 0.08% |

* Detailed solutions (code_and_data_matheuristic_for_SSCFLP_202105.zip) can be downloaded from https://github.com/yfkong/Unified.

## Table 7: Solution results on SSCFLP instances

| | CPLEX | | | | Matheuristic | | | | | |
|---|---|---|---|---|---|---|---|---|---|---|
| **Inst.** | LB | UB | Gap% | Time | Objmin | Objavg | Objmax | Gap | Stdev | Time |
| **zya1** | 4428.23 | 4428.23 | 0.00% | 877.53 | 4428.23 | 4428.23 | 4428.23 | 0.00% | 0.00% | 64.09 |
| **zya2** | 4722.23 | 4722.23 | 0.00% | 780.86 | 4722.23 | 4722.30 | 4722.57 | 0.00% | 0.00% | 60.08 |
| **zya3** | 5025.23 | 5025.23 | 0.00% | 988.81 | 5025.23 | 5025.97 | 5028.94 | 0.01% | 0.03% | 69.18 |
| **zya4** | 5327.60 | 5327.60 | 0.00% | 2042.05 | 5327.60 | 5327.87 | 5328.94 | 0.01% | 0.01% | 84.44 |
| **zya5** | 5626.94 | 5626.94 | 0.00% | 1458.30 | 5626.94 | 5627.41 | 5628.48 | 0.01% | 0.01% | 71.06 |
| **zyb1** | 3999.99 | 3999.99 | 0.00% | 6024.45 | 4006.75 | 4006.76 | 4006.80 | 0.17% | 0.00% | 57.28 |
| **zyb2** | 4239.99 | 4239.99 | 0.00% | 3368.16 | 4254.75 | 4254.76 | 4254.80 | 0.35% | 0.00% | 60.69 |
| **zyb3** | 4488.99 | 4488.99 | 0.00% | 7106.34 | 4511.75 | 4511.75 | 4511.75 | 0.51% | 0.00% | 77.15 |
| **zyb4** | 4736.99 | 4736.99 | 0.00% | 2676.81 | 4740.12 | 4762.83 | 4769.27 | 0.55% | 0.27% | 92.81 |
| **zyb5** | 4982.99 | 4982.99 | 0.00% | 2662.05 | 4982.99 | 4997.42 | 5018.16 | 0.29% | 0.37% | 88.09 |
| **zyc1** | 3732.85 | 3732.85 | 0.00% | 227.04 | 3732.85 | 3733.22 | 3733.77 | 0.01% | 0.01% | 89.21 |
| **zyc2** | 3940.85 | 3940.85 | 0.00% | 237.91 | 3940.85 | 3941.21 | 3941.77 | 0.01% | 0.01% | 73.03 |
| **zyc3** | 4158.85 | 4158.85 | 0.00% | 457.83 | 4158.85 | 4162.41 | 4175.31 | 0.09% | 0.17% | 95.08 |
| **zyc4** | 4374.85 | 4374.85 | 0.00% | 293.48 | 4374.85 | 4375.03 | 4375.74 | 0.00% | 0.01% | 86.59 |
| **zyc5** | 4589.85 | 4589.85 | 0.00% | 758.59 | 4589.85 | 4589.85 | 4589.85 | 0.00% | 0.00% | 102.85 |
| **gya1** | 3498793.22 | 3499012.55 | 0.01% | 7200.00 | 3499012.55 | 3499012.55 | 3499012.55 | 0.01% | 0.00% | 212.78 |
| **gya2** | 3657078.11 | 3657323.63 | 0.01% | 7200.00 | 3657299.50 | 3657343.39 | 3657442.02 | 0.01% | 0.00% | 241.03 |
| **gya3** | 3812527.81 | 3812605.72 | 0.00% | 7200.00 | 3812623.38 | 3812906.54 | 3813041.65 | 0.01% | 0.00% | 303.09 |
| **gya4** | 3966893.30 | 3966921.73 | 0.00% | 7200.00 | 3966941.43 | 3966967.68 | 3967072.69 | 0.00% | 0.00% | 277.41 |
| **gya5** | 4119130.39 | 4119543.12 | 0.01% | 7200.00 | 4119543.12 | 4120654.23 | 4125003.89 | 0.04% | 0.06% | 321.69 |
| **gyb1** | 3433138.33 | 3433223.99 | 0.00% | 7200.00 | 3433223.99 | 3433223.99 | 3433223.99 | 0.00% | 0.00% | 239.99 |
| **gyb2** | 3573531.57 | 3573616.42 | 0.00% | 7200.00 | 3573616.42 | 3573616.42 | 3573616.42 | 0.00% | 0.00% | 235.73 |
| **gyb3** | 3712631.49 | 3712713.37 | 0.00% | 7200.00 | 3712713.37 | 3712713.37 | 3712713.37 | 0.00% | 0.00% | 249.45 |
| **gyb4** | 3847355.36 | 3847386.80 | 0.00% | 7200.00 | 3847386.80 | 3847444.44 | 3847510.68 | 0.00% | 0.00% | 237.03 |
| **gyb5** | 3977990.12 | 3977990.12 | 0.00% | 2681.64 | 3978214.70 | 3978277.81 | 3978318.47 | 0.01% | 0.00% | 243.77 |
| **gyc1** | 3351292.23 | 3351292.23 | 0.00% | 587.87 | 3351292.23 | 3351434.45 | 3352003.36 | 0.00% | 0.01% | 188.27 |
| **gyc2** | 3475209.95 | 3475209.95 | 0.00% | 863.11 | 3475209.95 | 3475226.68 | 3475251.91 | 0.00% | 0.00% | 235.56 |
| **gyc3** | 3599127.67 | 3599127.67 | 0.00% | 1157.06 | 3599169.18 | 3599182.86 | 3599203.29 | 0.00% | 0.00% | 223.03 |
| **gyc4** | 3723045.39 | 3723045.39 | 0.00% | 137.38 | 3723045.39 | 3723486.60 | 3724189.24 | 0.01% | 0.02% | 270.69 |
| **gyc5** | 3840616.19 | 3840616.19 | 0.00% | 5198.12 | 3840663.00 | 3840721.03 | 3840746.99 | 0.00% | 0.00% | 260.66 |

## Table 8: Solution results on SSCKFLP instances

| | | CPLEX | | | | Matheuristic | | | | | |
|---|---|---|---|---|---|---|---|---|---|---|---|
| **Inst.** | K | LB | UB | Gap | Time | Objmin | Objavg | Objmax | Gap | Dev | Time |
| **ZYa1** | 13 | 4846.88 | 4846.88 | 0.00% | 140.14 | 4846.88 | 4850.45 | 4864.73 | 0.07% | 0.16% | 100.21 |
| **ZYa1** | 14 | 4740.14 | 4740.14 | 0.00% | 84.52 | 4745.55 | 4765.69 | 4785.13 | 0.54% | 0.39% | 85.56 |
| **ZYa1** | 15 | 4644.42 | 4644.42 | 0.00% | 26.14 | 4644.47 | 4644.47 | 4644.47 | 0.00% | 0.00% | 36.24 |
| **ZYa1** | 16 | 4549.83 | 4549.83 | 0.00% | 613.82 | 4557.03 | 4557.03 | 4557.03 | 0.16% | 0.00% | 65.49 |
| **ZYa1** | 17 | 4479.05 | 4479.05 | 0.00% | 7.91 | 4479.05 | 4479.05 | 4479.08 | 0.00% | 0.00% | 54.07 |
| **ZYa1** | 18 | 4459.38 | 4459.38 | 0.00% | 56.88 | 4459.38 | 4462.24 | 4463.34 | 0.06% | 0.04% | 62.40 |
| **ZYa1** | 19 | 4428.23 | 4428.23 | 0.00% | 585.82 | 4428.57 | 4433.88 | 4437.77 | 0.13% | 0.11% | 91.10 |
| **ZYa1** | 20 | 4441.19 | 4441.19 | 0.00% | 19.77 | 4441.19 | 4441.30 | 4441.74 | 0.00% | 0.01% | 42.14 |
| **ZYa1** | 21 | 4490.37 | 4490.37 | 0.00% | 4.83 | 4490.37 | 4490.37 | 4490.37 | 0.00% | 0.00% | 30.46 |
| **ZYa1** | 22 | 4617.61 | 4617.61 | 0.00% | 73.34 | 4617.88 | 4619.00 | 4623.07 | 0.03% | 0.05% | 27.89 |
| **zyb1** | 13 | 4162.76 | 4162.76 | 0.00% | 33.89 | 4162.76 | 4162.76 | 4162.76 | 0.00% | 0.00% | 46.41 |
| **zyb1** | 14 | 4104.85 | 4104.85 | 0.00% | 67.91 | 4104.85 | 4113.71 | 4124.17 | 0.22% | 0.23% | 81.31 |
| **zyb1** | 15 | 4061.12 | 4061.12 | 0.00% | 66.03 | 4061.12 | 4061.29 | 4061.87 | 0.00% | 0.01% | 72.98 |
| **zyb1** | 16 | 3999.99 | 3999.99 | 0.00% | 1243.20 | 4022.39 | 4025.35 | 4027.33 | 0.63% | 0.07% | 82.86 |
| **zyb1** | 17 | 4006.75 | 4006.75 | 0.00% | 23.85 | 4006.75 | 4006.75 | 4006.75 | 0.00% | 0.00% | 52.24 |
| **zyb1** | 18 | 4051.09 | 4051.09 | 0.00% | 21.99 | 4051.09 | 4051.19 | 4051.59 | 0.00% | 0.01% | 42.62 |
| **zyb1** | 19 | 4129.68 | 4129.68 | 0.00% | 20.44 | 4129.68 | 4129.69 | 4129.69 | 0.00% | 0.00% | 29.29 |
| **zyb1** | 20 | 4244.80 | 4244.80 | 0.00% | 10.77 | 4244.80 | 4244.80 | 4244.80 | 0.00% | 0.00% | 24.51 |
| **zyb1** | 21 | 4381.32 | 4381.32 | 0.00% | 11.99 | 4381.32 | 4381.32 | 4381.32 | 0.00% | 0.00% | 22.65 |
| **zyb1** | 22 | 4495.40 | 4495.40 | 0.00% | 7.63 | 4495.40 | 4495.40 | 4495.40 | 0.00% | 0.00% | 16.19 |
| **zyc1** | 13 | 3810.22 | 3810.22 | 0.00% | 70.63 | 3810.66 | 3812.22 | 3815.65 | 0.05% | 0.06% | 129.10 |
| **zyc1** | 14 | 3749.31 | 3749.31 | 0.00% | 47.81 | 3749.31 | 3752.38 | 3764.12 | 0.08% | 0.17% | 97.28 |
| **zyc1** | 15 | 3732.85 | 3732.85 | 0.00% | 54.00 | 3732.85 | 3733.22 | 3733.77 | 0.01% | 0.01% | 68.31 |
| **zyc1** | 16 | 3783.85 | 3783.85 | 0.00% | 38.72 | 3783.85 | 3783.85 | 3783.85 | 0.00% | 0.00% | 54.33 |
| **zyc1** | 17 | 3852.91 | 3852.91 | 0.00% | 15.84 | 3852.91 | 3852.91 | 3852.91 | 0.00% | 0.00% | 37.10 |
| **zyc1** | 18 | 3937.83 | 3937.83 | 0.00% | 12.91 | 3937.83 | 3937.99 | 3938.63 | 0.00% | 0.01% | 27.38 |
| **zyc1** | 19 | 4062.26 | 4062.26 | 0.00% | 8.64 | 4062.26 | 4062.26 | 4062.26 | 0.00% | 0.00% | 22.08 |
| **zyc1** | 20 | 4202.55 | 4202.55 | 0.00% | 11.57 | 4202.55 | 4202.55 | 4202.55 | 0.00% | 0.00% | 20.43 |
| **zyc1** | 21 | 4347.97 | 4347.97 | 0.00% | 9.39 | 4347.97 | 4347.97 | 4347.97 | 0.00% | 0.00% | 22.04 |
| **zyc1** | 22 | 4495.40 | 4495.40 | 0.00% | 7.07 | 4495.40 | 4495.40 | 4495.40 | 0.00% | 0.00% | 16.18 |
| **gya1** | 16 | 3653745.26 | 3654964.45 | 0.03% | 7290.77 | 3654964.45 | 3655006.87 | 3655092.78 | 0.03% | 0.00% | 323.50 |
| **gya1** | 17 | 3555195.99 | 3555687.10 | 0.01% | 7294.80 | 3555683.37 | 3555936.60 | 3556548.11 | 0.02% | 0.01% | 249.06 |
| **gya1** | 18 | 3516721.82 | 3517274.15 | 0.02% | 7213.19 | 3517195.85 | 3517293.37 | 3517521.50 | 0.02% | 0.00% | 252.80 |
| **gya1** | 19 | 3501647.12 | 3501975.32 | 0.01% | 7208.25 | 3502001.74 | 3504940.87 | 3509011.74 | 0.09% | 0.11% | 281.01 |
| **gya1** | 20 | 3498731.49 | 3499012.55 | 0.01% | 7206.58 | 3499012.55 | 3499018.31 | 3499028.92 | 0.01% | 0.00% | 230.72 |
| **gya1** | 21 | 3499745.27 | 3499745.27 | 0.00% | 122.78 | 3499745.27 | 3499749.25 | 3499751.90 | 0.00% | 0.00% | 185.58 |
| **gya1** | 22 | 3509618.91 | 3509618.91 | 0.00% | 63.27 | 3509618.91 | 3509622.42 | 3509629.85 | 0.00% | 0.00% | 123.21 |
| **gya1** | 23 | 3531467.24 | 3531467.24 | 0.00% | 50.42 | 3531467.24 | 3531467.24 | 3531467.24 | 0.00% | 0.00% | 101.74 |
| **gya1** | 24 | 3554134.39 | 3554134.39 | 0.00% | 40.91 | 3554134.39 | 3555665.84 | 3561744.78 | 0.04% | 0.10% | 85.21 |
| **gya1** | 25 | 3584373.45 | 3584373.45 | 0.00% | 15.19 | 3584380.08 | 3584390.60 | 3584421.26 | 0.00% | 0.00% | 81.31 |
| **gyb1** | 16 | 3457960.41 | 3457960.41 | 0.00% | 820.20 | 3457960.41 | 3457971.01 | 3458013.42 | 0.00% | 0.00% | 226.24 |
| **gyb1** | 17 | 3443416.95 | 3443416.95 | 0.00% | 1080.62 | 3443416.95 | 3443491.98 | 3443728.12 | 0.00% | 0.00% | 203.02 |
| **gyb1** | 18 | 3433223.99 | 3433223.99 | 0.00% | 578.38 | 3433223.99 | 3433223.99 | 3433223.99 | 0.00% | 0.00% | 216.82 |
| **gyb1** | 19 | 3440079.10 | 3440302.17 | 0.01% | 7280.07 | 3440302.17 | 3440304.39 | 3440313.28 | 0.01% | 0.00% | 229.67 |
| **gyb1** | 20 | 3451675.38 | 3451858.64 | 0.01% | 7389.10 | 3451858.64 | 3451858.64 | 3451858.64 | 0.01% | 0.00% | 238.56 |
| **gyb1** | 21 | 3468075.46 | 3468220.33 | 0.00% | 7298.66 | 3468220.33 | 3469381.89 | 3471035.89 | 0.04% | 0.03% | 206.84 |
| **gyb1** | 22 | 3481464.17 | 3481596.38 | 0.00% | 7319.51 | 3481596.38 | 3481756.83 | 3482398.64 | 0.01% | 0.01% | 230.82 |
| **gyb1** | 23 | 3503444.70 | 3503444.70 | 0.00% | 4738.72 | 3503444.70 | 3504750.89 | 3509903.22 | 0.04% | 0.08% | 160.05 |
| **gyb1** | 24 | 3531166.24 | 3531166.24 | 0.00% | 107.28 | 3531166.24 | 3531283.30 | 3531751.55 | 0.00% | 0.01% | 126.10 |
| **gyb1** | 25 | 3564169.79 | 3564169.79 | 0.00% | 81.95 | 3564169.79 | 3564329.43 | 3564568.88 | 0.00% | 0.01% | 105.51 |
| **gyc1** | 16 | 3395007.95 | 3395007.95 | 0.00% | 3343.99 | 3395021.33 | 3399827.50 | 3402802.51 | 0.14% | 0.12% | 282.67 |
| **gyc1** | 17 | 3369699.32 | 3369735.92 | 0.00% | 7386.79 | 3369747.59 | 3370087.37 | 3370332.64 | 0.01% | 0.01% | 234.17 |
| **gyc1** | 18 | 3353168.10 | 3353168.10 | 0.00% | 130.06 | 3353168.10 | 3353207.91 | 3353367.13 | 0.00% | 0.00% | 215.00 |
| **gyc1** | 19 | 3351292.23 | 3351292.23 | 0.00% | 136.91 | 3351292.23 | 3351687.55 | 3352219.88 | 0.01% | 0.01% | 206.98 |

## Table 8: Solution results on SSCKFLP instances (continued)

| | | CPLEX | | | | Matheuristic | | | | | |
|---|---|---|---|---|---|---|---|---|---|---|---|
| **Inst.** | K | LB | UB | Gap | Time | Objmin | Objavg | Objmax | Gap | Dev | Time |
| **gyc1** | 20 | 3364994.07 | 3364994.07 | 0.00% | 95.98 | 3364994.07 | 3364994.39 | 3364995.65 | 0.00% | 0.00% | 174.52 |
| **gyc1** | 21 | 3382762.01 | 3382762.01 | 0.00% | 88.54 | 3382762.01 | 3384394.16 | 3386842.40 | 0.05% | 0.07% | 162.15 |
| **gyc1** | 22 | 3404610.33 | 3404610.33 | 0.00% | 72.36 | 3404610.33 | 3409583.87 | 3414376.52 | 0.15% | 0.14% | 109.71 |
| **gyc1** | 23 | 3427712.02 | 3427712.02 | 0.00% | 39.84 | 3427712.02 | 3427712.65 | 3427713.60 | 0.00% | 0.00% | 111.79 |
| **gyc1** | 24 | 3459326.53 | 3459326.53 | 0.00% | 49.52 | 3459326.53 | 3459326.85 | 3459328.11 | 0.00% | 0.00% | 99.17 |
| **gyc1** | 25 | 3496498.72 | 3496498.72 | 0.00% | 44.11 | 3496498.72 | 3496498.72 | 3496498.72 | 0.00% | 0.00% | 75.00 |

## Table 9: Solution results on CFLSAP instances

| | CPLEX | | | | Matheuristic | | | | | |
|---|---|---|---|---|---|---|---|---|---|---|
| **Inst** | LB | UB | Gap | Time | Objmin | Objavg | Objmax | Gap | Dev | Time |
| **zya1** | 4435.21 | 4446.32 | 0.25% | 2h | 4446.79 | 4448.49 | 4451.36 | 0.30% | 0.04% | 66.76 |
| **zya2** | 4727.23 | 4740.69 | 0.28% | 2h | 4740.69 | 4741.91 | 4743.50 | 0.31% | 0.02% | 76.92 |
| **zya3** | 5031.18 | 5045.69 | 0.29% | 2h | 5045.69 | 5047.11 | 5050.89 | 0.32% | 0.04% | 86.97 |
| **zya4** | 5335.83 | 5344.66 | 0.17% | 2h | 5351.59 | 5354.74 | 5356.82 | 0.35% | 0.04% | 82.09 |
| **zya5** | 5634.87 | 5652.69 | 0.32% | 2h | 5653.06 | 5655.13 | 5658.20 | 0.36% | 0.05% | 74.15 |
| **zyb1** | 4014.58 | 4014.58 | 0.00% | 3034.54 | 4014.58 | 4015.45 | 4016.28 | 0.02% | 0.02% | 69.85 |
| **zyb2** | 4258.35 | 4263.26 | 0.12% | 2h | 4263.08 | 4263.55 | 4264.64 | 0.12% | 0.02% | 79.44 |
| **zyb3** | 4502.29 | 4519.58 | 0.38% | 2h | 4519.58 | 4520.58 | 4521.99 | 0.41% | 0.02% | 75.08 |
| **zyb4** | 4751.57 | 4778.24 | 0.56% | 2h | 4778.58 | 4779.57 | 4781.39 | 0.59% | 0.02% | 111.65 |
| **zyb5** | 4994.83 | 5032.58 | 0.75% | 2h | 5029.73 | 5033.72 | 5041.15 | 0.78% | 0.09% | 126.54 |
| **zyc1** | 3736.16 | 3736.16 | 0.00% | 5988.96 | 3736.44 | 3738.66 | 3742.82 | 0.07% | 0.08% | 61.10 |
| **zyc2** | 3941.78 | 3944.28 | 0.06% | 2h | 3945.37 | 3949.54 | 3952.72 | 0.20% | 0.09% | 73.86 |
| **zyc3** | 4161.52 | 4162.16 | 0.02% | 2h | 4163.21 | 4164.16 | 4165.99 | 0.06% | 0.03% | 64.39 |
| **zyc4** | 4372.17 | 4378.16 | 0.14% | 2h | 4378.44 | 4385.42 | 4390.15 | 0.30% | 0.10% | 65.94 |
| **zyc5** | 4593.16 | 4593.16 | 0.00% | 6698.44 | 4593.28 | 4594.53 | 4595.89 | 0.03% | 0.02% | 69.41 |
| **gya1** | 3494350.06 | 3499875.33 | 0.16% | 2h | 3499875.33 | 3499941.56 | 3500019.96 | 0.16% | 0.00% | 141.93 |
| **gya2** | 3656614.06 | 3658269.03 | 0.05% | 2h | 3658484.17 | 3660115.17 | 3661933.78 | 0.10% | 0.04% | 153.18 |
| **gya3** | 3811904.00 | 3814303.55 | 0.06% | 2h | 3813949.08 | 3814214.54 | 3814710.54 | 0.06% | 0.01% | 180.30 |
| **gya4** | 3950935.46 | 3968368.16 | 0.44% | 2h | 3968133.09 | 3968820.17 | 3969847.01 | 0.45% | 0.02% | 194.72 |
| **gya5** | 4111962.81 | 4121919.00 | 0.24% | 2h | 4120568.36 | 4121098.91 | 4121375.50 | 0.22% | 0.01% | 190.11 |
| **gyb1** | 3433407.72 | 3433407.72 | 0.00% | 5147.80 | 3433436.61 | 3433488.87 | 3433555.01 | 0.00% | 0.00% | 218.21 |
| **gyb2** | 3573800.15 | 3573800.15 | 0.00% | 6247.53 | 3573849.63 | 3573973.85 | 3574349.62 | 0.00% | 0.01% | 273.99 |
| **gyb3** | 3712194.67 | 3713059.77 | 0.02% | 2h | 3712958.80 | 3712996.46 | 3713039.84 | 0.02% | 0.00% | 265.43 |
| **gyb4** | 3844878.93 | 3852609.16 | 0.21% | 2h | 3847875.67 | 3848015.22 | 3848338.08 | 0.08% | 0.00% | 280.14 |
| **gyb5** | 3975431.45 | 3991801.41 | 0.41% | 2h | 3978777.75 | 3979011.83 | 3979277.87 | 0.09% | 0.00% | 245.70 |
| **gyc1** | 3353103.18 | 3355378.19 | 0.07% | 2h | 3356000.39 | 3356407.96 | 3357697.76 | 0.10% | 0.02% | 207.73 |
| **gyc2** | 3475414.69 | 3482023.47 | 0.19% | 2h | 3482090.41 | 3482274.51 | 3482500.26 | 0.20% | 0.00% | 214.37 |
| **gyc3** | 3599352.94 | 3612109.45 | 0.35% | 2h | 3605877.05 | 3606696.86 | 3608766.46 | 0.20% | 0.03% | 241.28 |
| **gyc4** | 3723704.34 | 3730327.88 | 0.17% | 2h | 3730810.21 | 3731823.15 | 3732412.06 | 0.22% | 0.02% | 240.87 |
| **gyc5** | 3839447.66 | 3852542.90 | 0.34% | 2h | 3850527.88 | 3852079.47 | 3855911.04 | 0.33% | 0.06% | 290.45 |

## Table 10: Solution results on CKFLSAP instances

| | | CPLEX | | | | Matheuristic | | | | | |
|---|---|---|---|---|---|---|---|---|---|---|---|
| Inst. | **K** | **LB** | **UB** | **Gap** | **Time** | **Objmin** | **Objavg** | **Objmax** | **Gap** | **Dev** | **Time** |
| ZYa1 | **13** | 4855.13 | 4855.13 | 0.00% | 4816.65 | 4856.05 | 4862.92 | 4873.44 | 0.16% | 0.15% | 62.67 |
| ZYa1 | **14** | 4745.68 | 4745.68 | 0.00% | 3474.80 | 4758.98 | 4793.57 | 4814.36 | 1.01% | 0.47% | 65.29 |
| ZYa1 | **15** | 4650.36 | 4650.36 | 0.00% | 2200.54 | 4652.08 | 4652.96 | 4654.00 | 0.06% | 0.02% | 49.96 |
| ZYa1 | **16** | 4562.21 | 4562.59 | 0.01% | 7222.15 | 4565.79 | 4571.13 | 4576.51 | 0.20% | 0.10% | 51.53 |
| ZYa1 | **17** | 4486.15 | 4486.15 | 0.00% | 2200.07 | 4486.21 | 4486.65 | 4486.85 | 0.01% | 0.01% | 58.28 |
| ZYa1 | **18** | 4469.29 | 4469.29 | 0.00% | 5032.55 | 4474.12 | 4477.42 | 4480.39 | 0.18% | 0.05% | 54.08 |
| ZYa1 | **19** | 4438.07 | 4441.13 | 0.07% | 7230.22 | 4470.36 | 4476.88 | 4487.35 | 0.87% | 0.15% | 79.63 |
| ZYa1 | **20** | 4446.32 | 4446.32 | 0.00% | 2078.38 | 4446.75 | 4450.49 | 4453.03 | 0.09% | 0.06% | 52.91 |
| ZYa1 | **21** | 4500.53 | 4500.53 | 0.00% | 394.76 | 4502.51 | 4506.63 | 4510.05 | 0.14% | 0.07% | 31.75 |
| ZYa1 | **22** | 4619.83 | 4620.24 | 0.01% | 7225.25 | 4622.12 | 4627.70 | 4634.33 | 0.17% | 0.10% | 33.33 |
| zyb1 | **13** | 4167.39 | 4167.39 | 0.00% | 5671.19 | 4168.25 | 4169.58 | 4171.64 | 0.05% | 0.04% | 46.74 |
| zyb1 | **14** | 4109.78 | 4109.78 | 0.00% | 2528.56 | 4111.40 | 4117.12 | 4126.68 | 0.18% | 0.15% | 63.84 |
| zyb1 | **15** | 4069.41 | 4069.41 | 0.00% | 6044.57 | 4079.84 | 4094.76 | 4100.28 | 0.62% | 0.21% | 66.68 |
| zyb1 | **16** | 4012.24 | 4036.50 | 0.60% | 7226.09 | 4036.68 | 4042.97 | 4060.01 | 0.77% | 0.24% | 54.39 |
| zyb1 | **17** | 4014.58 | 4014.58 | 0.00% | 842.52 | 4014.82 | 4015.37 | 4016.16 | 0.02% | 0.01% | 53.98 |
| zyb1 | **18** | 4059.26 | 4059.26 | 0.00% | 995.89 | 4061.27 | 4061.79 | 4062.10 | 0.06% | 0.01% | 37.52 |
| zyb1 | **19** | 4136.85 | 4136.85 | 0.00% | 318.25 | 4138.53 | 4140.21 | 4142.56 | 0.08% | 0.04% | 39.09 |
| zyb1 | **20** | 4250.06 | 4250.06 | 0.00% | 215.87 | 4250.06 | 4250.72 | 4251.75 | 0.02% | 0.02% | 27.37 |
| zyb1 | **21** | 4386.51 | 4386.51 | 0.00% | 188.30 | 4386.51 | 4386.78 | 4387.39 | 0.01% | 0.01% | 20.47 |
| zyb1 | **22** | 4525.63 | 4525.63 | 0.00% | 151.62 | 4525.63 | 4525.77 | 4525.89 | 0.00% | 0.00% | 16.52 |
| zyc1 | **13** | 3810.59 | 3810.59 | 0.00% | 1199.16 | 3811.21 | 3820.61 | 3830.46 | 0.26% | 0.23% | 61.42 |
| zyc1 | **14** | 3752.55 | 3752.55 | 0.00% | 537.32 | 3752.55 | 3758.94 | 3775.60 | 0.17% | 0.26% | 62.88 |
| zyc1 | **15** | 3736.16 | 3736.16 | 0.00% | 1354.40 | 3736.52 | 3740.80 | 3742.63 | 0.12% | 0.07% | 60.74 |
| zyc1 | **16** | 3785.44 | 3785.44 | 0.00% | 379.65 | 3786.57 | 3789.26 | 3791.77 | 0.10% | 0.07% | 34.32 |
| zyc1 | **17** | 3854.84 | 3854.84 | 0.00% | 245.93 | 3854.84 | 3855.91 | 3856.34 | 0.03% | 0.02% | 30.56 |
| zyc1 | **18** | 3939.11 | 3939.11 | 0.00% | 191.12 | 3939.11 | 3939.28 | 3939.72 | 0.00% | 0.01% | 29.00 |
| zyc1 | **19** | 4062.63 | 4062.63 | 0.00% | 162.81 | 4062.63 | 4062.63 | 4062.63 | 0.00% | 0.00% | 22.15 |
| zyc1 | **20** | 4202.91 | 4202.91 | 0.00% | 156.69 | 4202.91 | 4205.10 | 4205.64 | 0.05% | 0.03% | 21.70 |
| zyc1 | **21** | 4348.33 | 4348.33 | 0.00% | 158.86 | 4348.33 | 4349.50 | 4354.19 | 0.03% | 0.06% | 16.88 |
| zyc1 | **22** | 4495.76 | 4495.76 | 0.00% | 150.90 | 4495.76 | 4495.76 | 4495.76 | 0.00% | 0.00% | 15.12 |
| gya1 | **16** | 3654122.32 | 3655741.03 | 0.03% | 7280.55 | 3656580.12 | 3657435.11 | 3660149.56 | 0.09% | 0.04% | 140.15 |
| gya1 | **17** | 3554253.35 | 3557622.55 | 0.09% | 7296.34 | 3558134.69 | 3558726.50 | 3559890.85 | 0.13% | 0.02% | 136.55 |
| gya1 | **18** | 3513204.32 | 3522650.84 | 0.27% | 7260.45 | 3518604.91 | 3519707.77 | 3522815.57 | 0.19% | 0.05% | 130.35 |
| gya1 | **19** | 3501266.20 | 3509894.00 | 0.25% | 7262.72 | 3503706.90 | 3506974.94 | 3509115.11 | 0.16% | 0.08% | 168.43 |
| gya1 | **20** | 3498654.87 | 3500110.57 | 0.04% | 7253.28 | 3499833.18 | 3500281.17 | 3500930.65 | 0.05% | 0.01% | 125.11 |
| gya1 | **21** | 3500566.31 | 3500566.31 | 0.00% | 4194.36 | 3500566.31 | 3500590.95 | 3500649.50 | 0.00% | 0.00% | 143.15 |
| gya1 | **22** | 3510439.96 | 3510439.96 | 0.00% | 4944.07 | 3510439.96 | 3510489.36 | 3510613.95 | 0.00% | 0.00% | 97.40 |
| gya1 | **23** | 3532049.19 | 3532374.32 | 0.01% | 7248.48 | 3532288.28 | 3532331.32 | 3532407.38 | 0.01% | 0.00% | 102.63 |
| gya1 | **24** | 3554955.43 | 3554955.43 | 0.00% | 1935.15 | 3554955.43 | 3554968.53 | 3555012.30 | 0.00% | 0.00% | 89.50 |
| gya1 | **25** | 3585194.49 | 3585194.49 | 0.00% | 803.02 | 3585225.78 | 3585302.88 | 3585446.72 | 0.00% | 0.00% | 72.14 |
| gyb1 | **16** | 3457960.41 | 3457960.41 | 0.00% | 3674.65 | 3457960.41 | 3458044.90 | 3458116.48 | 0.00% | 0.00% | 175.33 |
| gyb1 | **17** | 3443566.20 | 3443654.54 | 0.00% | 7310.09 | 3443648.14 | 3443790.77 | 3444004.15 | 0.01% | 0.01% | 218.20 |
| gyb1 | **18** | 3433407.72 | 3433407.72 | 0.00% | 3416.07 | 3433407.72 | 3433462.27 | 3433544.08 | 0.00% | 0.00% | 185.56 |
| gyb1 | **19** | 3440483.00 | 3440485.90 | 0.00% | 7280.34 | 3440505.76 | 3440625.47 | 3440964.83 | 0.00% | 0.01% | 150.67 |
| gyb1 | **20** | 3452042.37 | 3452042.37 | 0.00% | 5666.14 | 3452042.37 | 3452081.54 | 3452145.18 | 0.00% | 0.00% | 147.24 |
| gyb1 | **21** | 3468404.06 | 3468404.06 | 0.00% | 3196.31 | 3468423.91 | 3469203.01 | 3472224.24 | 0.02% | 0.05% | 129.04 |
| gyb1 | **22** | 3481720.97 | 3489079.64 | 0.21% | 7310.84 | 3489079.64 | 3489355.58 | 3490335.50 | 0.22% | 0.02% | 122.27 |
| gyb1 | **23** | 3505040.26 | 3512497.85 | 0.21% | 7293.16 | 3510961.78 | 3514895.20 | 3516406.48 | 0.28% | 0.06% | 131.56 |
| gyb1 | **24** | 3531221.55 | 3538649.50 | 0.21% | 7281.04 | 3538649.50 | 3539589.67 | 3543350.38 | 0.24% | 0.06% | 101.09 |
| gyb1 | **25** | 3564485.99 | 3572451.06 | 0.22% | 7261.95 | 3572451.06 | 3572451.06 | 3572451.06 | 0.22% | 0.00% | 89.48 |
| gyc1 | **16** | 3398707.42 | 3410837.65 | 0.36% | 7285.12 | 3400531.44 | 3409706.70 | 3418630.56 | 0.32% | 0.19% | 284.12 |
| gyc1 | **17** | 3374018.35 | 3375232.53 | 0.04% | 5268.97 | 3375317.09 | 3375583.62 | 3376280.81 | 0.05% | 0.01% | 211.47 |
| gyc1 | **18** | 3357023.41 | 3358443.09 | 0.04% | 7294.38 | 3358390.63 | 3359053.33 | 3359907.80 | 0.06% | 0.02% | 184.26 |
| gyc1 | **19** | 3353441.88 | 3358535.61 | 0.15% | 7299.93 | 3356072.72 | 3357079.34 | 3357832.46 | 0.11% | 0.03% | 164.07 |

## Table 10: Solution results on CKFLSAP instances (continued)

| | | CPLEX | | | | Matheuristic | | | | | |
|---|---|---|---|---|---|---|---|---|---|---|---|
| Inst. | **K** | **LB** | **UB** | **Gap** | **Time** | **Objmin** | **Objavg** | **Objmax** | **Gap** | **Dev** | **Time** |
| gyc1 | **20** | 3366845.03 | 3366845.03 | 0.00% | 4784.34 | 3367891.52 | 3368513.01 | 3369364.37 | 0.05% | 0.02% | 134.12 |
| gyc1 | **21** | 3384608.97 | 3384608.97 | 0.00% | 2947.04 | 3385765.32 | 3389025.08 | 3391126.97 | 0.13% | 0.07% | 112.44 |
| gyc1 | **22** | 3406457.30 | 3406457.30 | 0.00% | 6192.43 | 3406461.84 | 3408993.58 | 3411589.85 | 0.07% | 0.05% | 93.97 |
| gyc1 | **23** | 3429558.99 | 3429558.99 | 0.00% | 1355.10 | 3430605.47 | 3431729.81 | 3432221.77 | 0.06% | 0.02% | 77.56 |
| gyc1 | **24** | 3461173.50 | 3461173.50 | 0.00% | 1617.16 | 3462498.58 | 3463401.87 | 3464482.69 | 0.06% | 0.02% | 71.01 |
| gyc1 | **25** | 3498357.39 | 3498357.39 | 0.00% | 1538.64 | 3499483.41 | 3500267.37 | 3501018.41 | 0.05% | 0.02% | 63.54 |